\documentclass[
    twocolumn, 
    amsmath,
    amssymb,
    aps,
    prl,
    showkeys,
    showpacs,
    floatfix,
    superscriptaddress,
    preprintnumbers] {revtex4-2}

\usepackage[utf8]{inputenc}
\usepackage{hyperref}
\usepackage{graphicx}
\usepackage{dcolumn}
\usepackage{bm}
\usepackage{tikz}
\usepackage[pdftex, dvipsnames]{xcolor}
\usepackage{xargs}
\usepackage{lipsum}
\usepackage{soul}
\usepackage{booktabs}
\usepackage{comment}
\usepackage{braket}
\usepackage{slashed}
\usetikzlibrary{external}
\usetikzlibrary{calc}
\usetikzlibrary{arrows.meta}
\usetikzlibrary{backgrounds}
\usetikzlibrary{intersections}
\usetikzlibrary{fit}
\usetikzlibrary{arrows.meta}
\usetikzlibrary{matrix}
\usetikzlibrary{shapes}
\usetikzlibrary{3d}
\pgfdeclarelayer{background}
\pgfdeclarelayer{middle}
\pgfdeclarelayer{foreground}
\pgfsetlayers{background,middle,main,foreground}
\usepackage[colorinlistoftodos,prependcaption,textsize=tiny]{todonotes}

\newcommandx{\unsure}[2][1=]{\todo[linecolor=red,backgroundcolor=red!25,bordercolor=red,#1]{#2}}
\newcommandx{\change}[2][1=]{\todo[linecolor=blue,backgroundcolor=blue!25,bordercolor=blue,#1]{#2}}
\newcommandx{\info}[2][1=]{\todo[linecolor=OliveGreen,backgroundcolor=OliveGreen!25,bordercolor=OliveGreen,#1]{#2}}
\newcommandx{\improvement}[2][1=]{\todo[linecolor=Plum,backgroundcolor=Plum!25,bordercolor=Plum,#1]{#2}}
\newcommandx{\thiswillnotshow}[2][1=]{\todo[disable,#1]{#2}}

\definecolor{mplMaroon}{HTML}{800000}      
\definecolor{mplForestGreen}{HTML}{228B22} 
\definecolor{mplPurple}{HTML}{800080}      

\newcommand{\nn}{\nonumber \\}

\begin{document}

\title{Simulating Lattice Gauge Theories with Virtual Rishons}

\author{David Rogerson}
 \email{dmr369@physics.rutgers.edu}
\affiliation{Department of Physics and Astronomy, Rutgers University, Piscataway, NJ 08854-8019 USA}
\author{João Barata}
\affiliation{CERN, Theoretical Physics Department, CH-1211, Geneva 23, Switzerland}
\author{Robert M. Konik}
\affiliation{Division of Condensed Matter Physics and Material Science, Brookhaven National Laboratory, Upton, NY 11973-5000, USA}

\author{Raju Venugopalan}
\affiliation{Physics Department, Brookhaven National Laboratory, Upton, New York 11973, USA} 
\affiliation{CFNS, Department of Physics and Astronomy, Stony Brook University, Stony Brook, NY 11794, USA}
\affiliation{Higgs Center for Theoretical Physics, The University of Edinburgh, Edinburgh, EH9 3FD, Scotland, UK}

\author{Ananda Roy}
\email{ananda.roy@physics.rutgers.edu}
\affiliation{Department of Physics and Astronomy, Rutgers University, Piscataway, NJ 08854-8019 USA}

\preprint{CERN-TH-2026-011}

\begin{abstract}
 Classical tensor network and hybrid quantum-classical algorithms are 
 promising candidates for the investigation of real-time properties of lattice gauge theories.We develop here a novel framework which enforces gauge symmetry via a quantum-link \textit{virtual rishon} representation applied at intermediate steps. Crucially, the gauge and matter degrees of freedom are dynamical variables encoded in terms of qubits, enabling analysis of gauge theories in~$d+1$ spacetime dimensions. We benchmark this framework in a U(1) gauge theory with and without matter fields. For $d = 1$, the multi-flavor Schwinger model with~$1\leq N_f\leq3$ flavors is analyzed for arbitrary boundary conditions and nonzero topological angle, capturing signatures of the underlying Wess-Zumino-Witten conformal field theory. For $d = 2$, we extract the confining string tension in close agreement with continuum expectations. These results establish the virtual rishon framework as a scalable and robust approach for the simulation of lattice gauge theories using both classical tensor networks as well as near-term quantum hardware. 
\end{abstract}

\maketitle

\emph{Introduction.} An ab initio understanding of the real-time dynamics of quantum field theories remains elusive. This is of particular relevance in regimes where the dynamics is intrinsically non-perturbative and not amenable to Euclidean Monte Carlo methods~\cite{Saleur:1998hq,annurev:/content/journals/10.1146/annurev-conmatphys-031620-102024,Alexandru:2020wrj,Berges:2020fwq,Eisert:2014jea}. 
Recent advances in quantum information science (QIS) --- notably tensor-network algorithms and programmable quantum hardware --- allow one, in certain contexts, to bypass this problem and simulate gauge theories directly in real time~\cite{Banuls:2018jag,Banuls:2019bmf,Magnifico:2024eiy,Bauer:2022hpo,Halimeh:2025vvp,Bauer:2023qgm}. 
In particular, over the past decade there has been substantial progress on studying real-time dynamics in low-dimensional gauge theories, including particle production~\cite{Schmidt:2024zpg,Maceda:2024rrd,Budd:2026rmw}, string dynamics~\cite{Gonzalez-Cuadra:2024xul,Cobos:2025krn,Xu:2025abo,Tian:2025mbv,Cataldi:2025cyo,Barata:2023jgd,Artiaco:2025qqq,bombieri2026,Cochran2025,xu2025}, scattering~\cite{Pavesic:2025nwm,Rigobello:2021fxw,Davoudi:2024wyv,Barata:2020jtq,Jordan:2011ci,Schuhmacher:2025ehh,Joshi:2025rha,Farrell:2025nkx,Zemlevskiy:2024vxt,Hardy:2024ric,Barata:2025hgx,Barata:2024apg,Barata:2025rjb,papaefstathiouRealtimeScatteringLattice2025,Barata:2023clv}, and thermalization~\cite{Qian:2024xnr,Ikeda:2024rzv,Desaules:2022ibp,Desaules:2022kse,Luo:2023rso,Banuls:2010zki,Zhou:2021kdl,Barata:2025jhd,Angelides:2025hjt,Chen:2024pee}.

Despite these developments, extending QIS approaches to higher dimensional models, more complex gauge groups, and richer matter content theories remains challenging~\cite{Raychowdhury:2018osk,Anishetty:2009nh,Raychowdhury:2019iki,Ilcic:2026cac,Ciavarella:2024lsp,Ciavarella:2025bsg,Grabowska:2024emw,Gustafson:2024kym,Carena:2022kpg,Kadam:2025trs,Kadam:2024ifg,Chandrasekharan:1996ih,Brower1999,Ale:2025sxz,Crane:2024tlj,PRXQuantum.5.040309,Halimeh2025,Cochran2025,PhysRevD.110.014507,PhysRevD.110.054511,PhysRevLett.133.111901,wu2025,bombieri2026,Cochran2025,xu2025}. These difficulties are intimately tied to the presence of local constraints in gauge theories that restrict the physical Hilbert space. In lattice formulations, faithfully enforcing these constraints requires large effective local Hilbert spaces, which grow rapidly with the gauge group representation, spatial dimension, and number of matter fields. This severely limits both classical tensor network simulations and implementations on quantum hardware, where resource availability is limited. Simulations of gauge theories relevant for applications therefore require the development of resource-efficient scalable formulations preserving gauge symmetry.

\begin{figure}
    \centering
    \includegraphics[width=.9\linewidth]{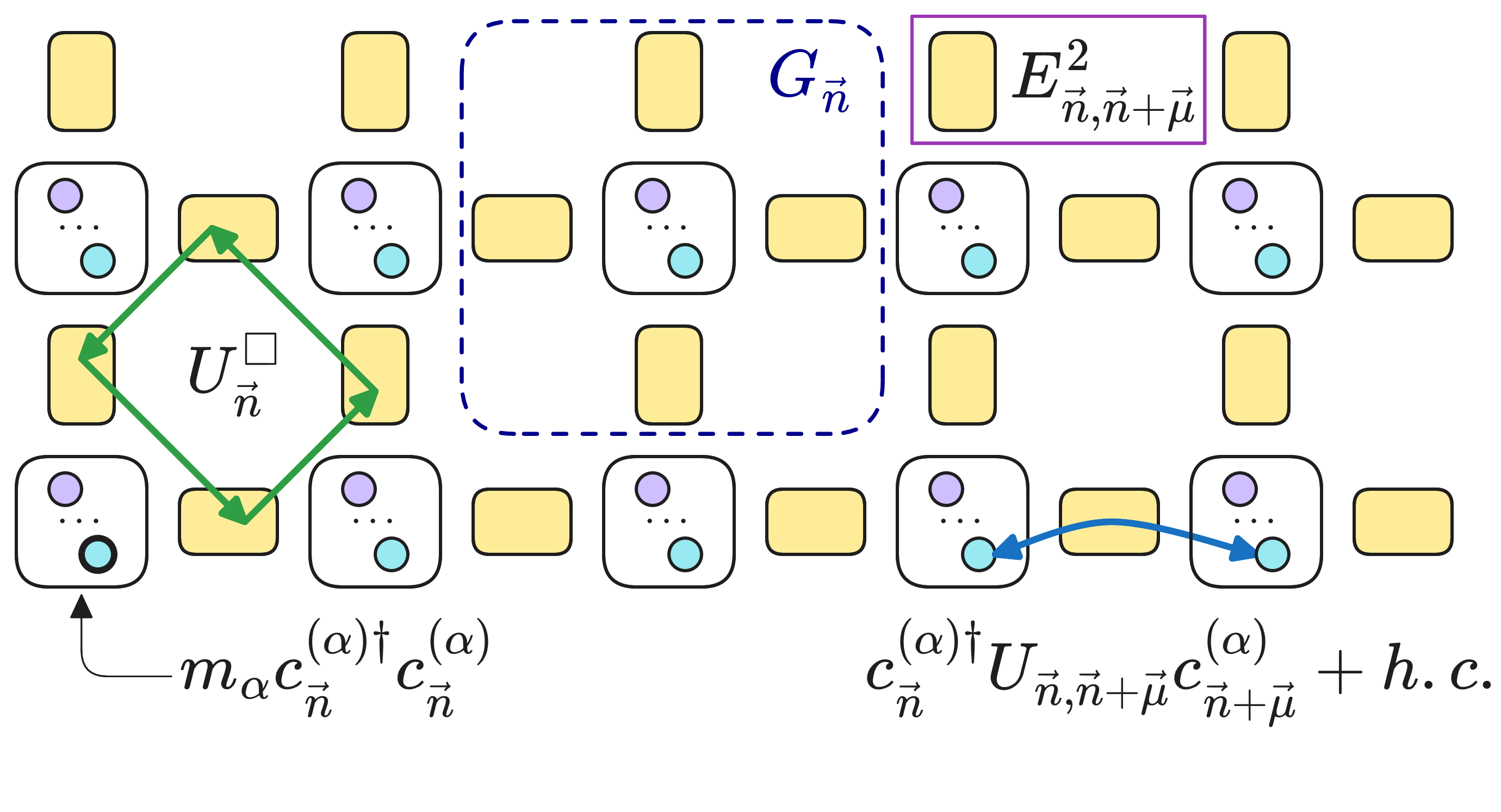}
    \vspace{-2em}
    \caption{Schematic of interactions in a U(1) lattice gauge theory. Fermion matter degrees of freedom (colored circles) sit on lattice sites connected by gauge links (yellow rectangles). The contributions to the Hamiltonian~[Eq.~\eqref{eq:Hamiltonian}] and the support of the local Gauss operator $G_{\vec n}$~[Eq.~\eqref{eq:gausslaw}] are indicated.}
    \label{fig:LatticeSchematic}
\end{figure}

We address this challenge by developing a novel virtual rishon (VR) framework amenable for analysis of lattice gauge theory models using  classical tensor networks and quantum simulation protocols.  A key feature is the quantum-link rishon representation~\cite{Chandrasekharan:1996ih,Brower1999} applied at intermediate steps to analytically identify non-overlapping local gauge charge. This allows gauge invariance to be encoded exactly while separating gauge and matter degrees of freedom. Crucially, in this VR formulation, the rishons serve as an auxiliary analytical tool to infer symmetry-preserving parametrizations but do not enlarge the physical Hilbert space of the theory. As we will elaborate, this enables the construction of symmetry-conserving tensor-network representations and qubit encodings with significantly reduced local resource requirements.

\emph{Virtual rishon construction of $U(1)$ lattice gauge theory.}
\begin{figure*}
    \centering
    \includegraphics[width=0.98\textwidth]{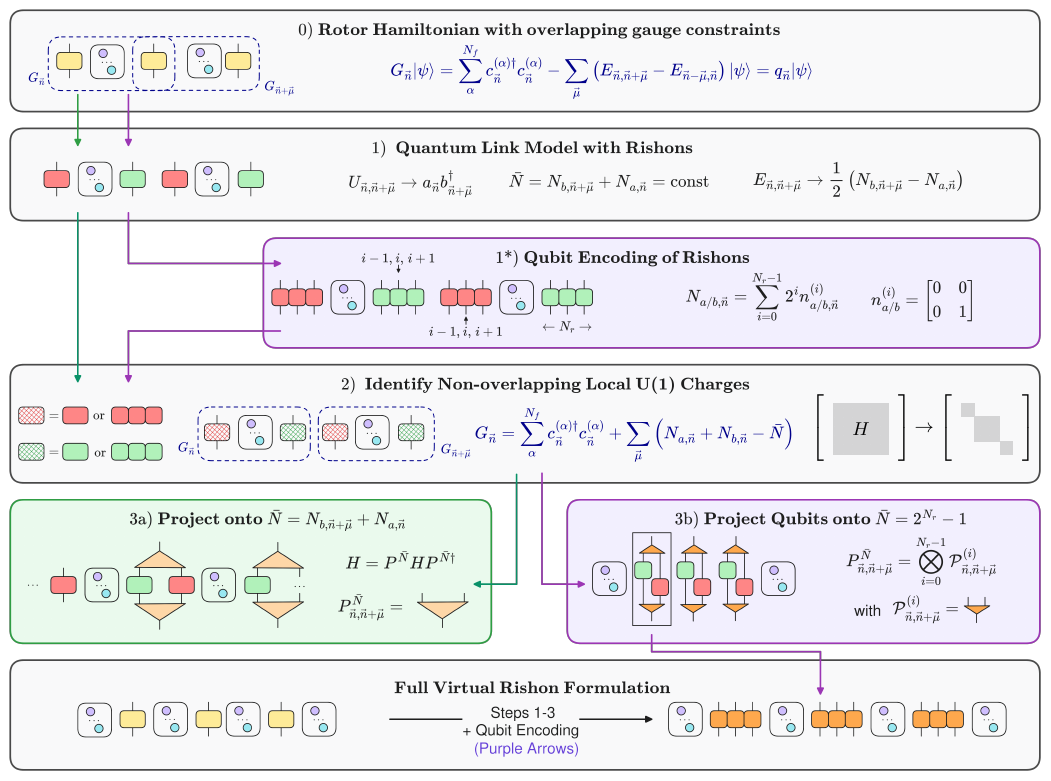}
    \vspace{-1em}
    \caption{\label{fig:RishonConstruction} Virtual rishon formulation. Green arrows indicate the generic formulation, while purple arrows include the qubit encoding.
           0) Hamiltonian with matter fields on the sites and U(1) rotors on the links; the Hilbert space is restricted due to the overlapping gauge constraints (dark blue boxes). 1) Each rotor is represented by two bosonic rishons (red and green boxes). 1*) Binary encoding of rishons using $N_r$-qubits.
           2) In the rishon representation, Gauss law constraints do not overlap and the Hamiltonian block-diagonalizes into irreducible representations of the gauge group.
           3a) The rishon link constraint $\bar{N}=N_a+N_b$ is enforced by projecting $H$ onto the relevant subspace via $P^{\bar{N}}$. 3b) In the qubit encoding, $P^{\bar{N}}$ factorizes into a product of two-qubit site projectors $\mathcal{P}^{(i)}$.
           The projector $P^{\bar{N}}$ preserves the block-diagonal structure of the Hamiltonian, yielding the qubit representation that explicitly conserves Gauss's law.
           }
\end{figure*}
We consider the Kogut-Susskind Hamiltonian~\cite{Kogut:1974ag,Susskind:1976jm} of U(1) lattice gauge theory in $d$ spatial dimensions with $N_f$ fundamental fermions. In the temporal gauge, the theory maps to the lattice model:
\begin{align}
H &= \frac{g^2a}{2}\sum_{\vec{n},\vec{\mu}} E_{\vec{n},\vec{n} + \vec{\mu}}^2 + \sum_{\vec{n}} \sum_{\alpha=1}^{N_{f}} (-1)^{\vec{n}} m_{\alpha} c^{(\alpha) \dagger}_{\vec{n}}c^{(\alpha)}_{\vec{n}} \nn 
    &\quad+\frac{1}{2a}\sum_{\vec{n}, \vec{\mu}} \sum_{\alpha=1}^{N_{f}} \eta_{\vec \mu}(\vec n)\left(c^{\dagger(\alpha)}_{\vec{n}} U_{\vec{n}, \vec{n} + \vec{\mu}} 
    c^{(\alpha)}_{\vec{n}+ \vec{\mu}} + h.c. \right) \nn
    &\quad- \frac{1}{2 g^2 a}\sum_{\vec{n}, \vec{\mu}_1 \neq \vec{\mu}_2} \left(U^{\Box}_{\vec{n}} + U^{\Box \dagger} _{\vec{n}} \right) \, , \label{eq:Hamiltonian}
\end{align}
graphically summarized in Fig.~\ref{fig:LatticeSchematic}; we work in dimensionless units obtained by rescaling the Hamiltonian. 
The matter content is described by 
fermionic operators $c^{(\alpha)}_{\vec{n}}$ with flavor-dependent mass $m_\alpha$; $\eta_{\vec \mu}(\vec n)$ is the staggered lattice gamma matrix remnant~\cite{Rothe:1992nt}.
The electric field operator $E_{\vec{n}, \vec{n} + \vec{\mu}}$ acts on links connecting staggered fermionic sites, 
with lattice spacing $a$. It satisfies the commutation relation $\left[E_{\vec{n}, \vec{n} + \vec{\mu}}, U_{\vec{m}, \vec{m} + \vec{\mu}_2}\right] = \delta_{\vec{n},\vec{m}}\delta_{\vec{\mu},\vec{\mu}_2}U_{\vec{n}, \vec{n} + \vec{\mu}}$, with $U_{\vec{n}, \vec{n} + \vec{\mu}} = \exp(i \phi_{\vec{n}, \vec{n} + \vec{\mu}})$ the link operator written in terms of $\phi$, the operator conjugate to $E$.  The plaquette operator $U^{\Box}_{\vec{n}} =  U_{\vec{n}, \vec{n} + \vec{\mu}_1}U_{\vec{n} + \vec{\mu}_1, \vec{n} + \vec{\mu}_1 + \vec{\mu}_2}U^\dagger_{\vec{n} + \vec{\mu}_2, \vec{n} + \vec{\mu}_1 + \vec{\mu}_2}U^\dagger_{\vec{n}, \vec{n} + \vec{\mu}_2}$ corresponds to the energy of the magnetic field in $d\ge 2$.
In the Hamiltonian formalism, gauge invariance is ensured by restricting to states $|\psi \rangle$  locally satisfying Gauss's Law, $G_{\vec{n}}\ket{\psi} = q_{\vec{n}}\ket{\psi}$, with the Gauss operator
\begin{align}
G_{\vec{n}}  &=  \sum_\alpha  c^{(\alpha) \dagger}_{\vec{n}} c^{(\alpha)}_{\vec{n}} - \sum_{\vec{\mu}}\left(E_{\vec{n}, \vec{n} +  \vec{\mu}} - E_{\vec{n} -  \vec{\mu}, \vec{n}} \right)  
 \, . \label{eq:gausslaw}
\end{align}

The physical Hilbert space decomposes into superselection sectors labeled by the abelian charges $q_{\vec n}$. As the local Gauss operators commute with each other, and with the Hamiltonian, the eigenspace factorizes into independent sectors corresponding to different background charge distributions. Restricting the dynamics to one sector guarantees gauge invariance; see the Supplemental Material (SM) for more details.

 The conservation of the U(1) gauge symmetry is also crucial 
 for predictions of physical quantities. This is true for both classical and quantum algorithms. Indeed, approaches that explicitly respect gauge symmetry suitable for implementation in tensor networks and quantum hardware have been developed~\cite{silviLatticeGaugeTensor2014, buyensMatrixProductStates2014, canalsTensorNetworkFormulation2024}. However, they require the enlargement of the effective local Hilbert space,
 combining both gauge and matter degrees of freedom. This results in higher computational resource requirements and the emergence of nonlocal interactions, particularly when considering multi-flavor models or extensions to non-abelian groups.

 As described below, the VR formalism avoids these shortcomings by separating gauge and matter degrees of freedom. This allows for a direct map to the original lattice model, explicitly building in gauge invariance by employing a non-overlapping construction of local conserved charges. Of particular importance 
 is the resulting qubit encoding of both  gauge and matter degrees of freedom.
 The steps in the VR construction are shown in Fig~\ref{fig:RishonConstruction} both for a generic construction (following green arrows) and the qubit encoded variant (purple arrows). 

The first step of the VR construction in  Fig~\ref{fig:RishonConstruction} is based on the quantum link model~\cite{Brower1999, wieseUltracoldQuantumGases2013, silviLatticeGaugeTensor2014}; here we express the gauge link operator as pairs of bosonic rishons $\{b_{\vec{n}}^{(\mu)}, a_{\vec{n}}^{(\mu)}\}$, in contrast to fermionic pairs in the original formulation. The single subscript emphasizes that the rishons are site rather than link centric; the superscript $\mu$ labels the spatial direction. They are related to link and electric operators via the standard map~\cite{Chandrasekharan:1996ih}: 
\begin{align}
    U_{\vec{n},\vec{n} + \vec{\mu}} \to a^{(\mu)}_{\vec{n}}b^{(\mu)\dagger}_{\vec{n}+\vec{\mu}} \,\,\,\! ;\! \,\,\,\!\! E_{\vec{n}, \vec{n} + \vec{\mu}} \to \frac{N^{(\mu)}_{b,\vec{n} + \vec{\mu}} - N^{(\mu)}_{a,\vec{n}}}{2}\, ,
\end{align}
where $N^{(\mu)}_{a/b}$ is the bosonic number operator for a given species. This representation is equivalent to the original model with a truncated rotor of dimension $\bar{N}+1$, while conserving the link constraints $\bar{N} = N^{(\mu)}_{a,\vec{n}} + N^{(\mu)}_{b,\vec{n} + \vec{\mu}}$, limited to maximal occupancy $\mathrm{Dim}(N_{a/b}^{(\mu)}) = \bar{N}$. Following the generic construction (green arrow) to step  2, the Gauss operator reads 
\begin{align}
    G_{\vec{n}} =  \sum_\alpha^{N_{f}} c^{(\alpha) \dagger}_{\vec{n}}c^{(\alpha)}_{\vec{n}} + \sum_{\vec{\mu}}\left(N^{(\mu)}_{a,\vec{n}} +N^{(\mu)}_{b,\vec{n}} - \bar{N}\right)\,, \label{eq:gaugeOpRishon}
\end{align}
which now has strictly local support. Each interaction term in $H$ is a tensor product of site-local operators $O_{\vec{n}}$. Crucially, each $O_{\vec{n}}$ now transforms under a single local charge $G_{\vec{n}}$, which in turn allows the global interaction to be parametrized via the irreducible representations of the individual sites (details in the SM). This extensive set of local conserved quantities reduces computational complexity while preserving Gauss's law.

As indicated in Fig~\ref{fig:RishonConstruction} (step~$3a$), the newly introduced link constraint on the combined bosonic occupation numbers is enforced by projecting on to the relevant subspace: 
\begin{align}
P^{\bar{N}}_{\vec{n}, \vec{n} + \vec{\mu}}\! =\!\sum_{e = -\bar{N}/2}^{\bar{N}/2} \ket{e}_{\vec{n}, \vec{n} + \vec{\mu}}\bra{\bar{N}/2-e}_{a, \vec{n}}^{(\mu)}\bra{\bar{N}/2+e}^{(\mu)}_{b, \vec{n} + \vec{\mu}} \, ,
\end{align}
where $\{ \ket{e} \}$ labels the eigenbasis of the electric field operator and 
the projector fuses the rishon pair on each link back into a single rotor degree of freedom. 
This approach differs qualitatively from treatments employing energy penalties~\cite{Banerjee:2012pg,Zohar:2015hwa}, parametrization in terms of irreducible representations~\cite{silviLatticeGaugeTensor2014,Tagliacozzo:2012df}, using gauge redundancy as resource for quantum error correction~\cite{PhysRevD.110.054516,pato2026}, or exploiting measurements to enforce gauge invariance~\cite{PhysRevB.111.094315}.

Operating with the projector on the Hamiltonian (employing $
\tilde{U}_{\vec{n}, \vec{n} + \vec{\mu}} =  P_{\vec{n}, \vec{n} + \vec{\mu}} a^{(\mu)}_{\vec{n}}b^{(\mu)\dagger}_{\vec{n}+\vec{\mu}} P^\dagger_{\vec{n}, \vec{n} + \vec{\mu}}$), 
the initial Hamiltonian is recovered with local Hilbert space dimensions identical to the initial problem. Importantly, the local gauge conserving Gauss law representation  carries forward even after the projection: the VR quantum link formulation is only used as an \textit{intermediate} bookkeeping step.
This procedure treats local matter and gauge Hilbert spaces separately, a key novel feature of the VR formulation relative to other approaches~\cite{wieseUltracoldQuantumGases2013,silviLatticeGaugeTensor2014, buyensMatrixProductStates2014}. It is ideal for simulating physical models with complex matter content.

The VR framework  
can be developed as a qubit-level system relevant for implementation on quantum hardware. This is shown by the purple path of Fig~\ref{fig:RishonConstruction} (step  1*). Each rishon number operator is first represented by a binary expansion using $N_r$ qubits:
\begin{align}
&N^{(\mu)}_{a/b,\vec{n}} = \sum_{i=0}^{N_r- 1} 2^i n^{(i,\mu)}_{a/b,\vec{n}} \, ,
\end{align}
where the number operator $n^{(i,\mu)}_{a,\vec{n}} = \frac{1}{2}(\mathbb{I}- Z^{(i,\mu)}_{a,\vec{n}})$ is represented by a Pauli $Z$ operator, with $(i)$ labeling the binary power. While the block-diagonal representation of the Hamiltonian is found analogously to the generic formulation (step 2), the projection (step 3b) is especially efficient for maximal encoding density by setting $\bar{N} = 2^{N_r}-1$. The $2N_r$--qubit projector $P_{\vec{n}, \vec{n} + \vec{\mu}}^{\bar{N}} = \bigotimes_{i=0}^{N_r -1} \mathcal{P}^{(i)}_{\vec{n}, \vec{n} + \vec{\mu}}$ factorizes into projectors on the odd parity sector of  equal binary power rishon-qubit pairs:
\begin{align}
\mathcal{P}^{(i)}_{\vec{n}, \vec{n} + \vec{\mu}} &= \ket{\uparrow}^{(i)}_{\vec{n}, \vec{n} + \vec{\mu}}   \bra{0}^{(i,\mu)}_{a,\vec{n}} \bra{1}^{(i,\mu)}_{b,\vec{n} + \vec{\mu}} \nn &+ \ket{\downarrow}^{(i)}_{\vec{n}, \vec{n} + \vec{\mu}}   \bra{1}^{(i,\mu)}_{a,\vec{n}} \bra{0}^{(i,\mu)}_{b,\vec{n} + \vec{\mu}}.
\end{align}
The labels of the restricted eigenbasis $\ket{\uparrow/\downarrow}_{\vec{n}, \vec{n} + \vec{\mu}}$ 
are chosen such that they contribute positively or negatively towards the total  integer valued electric field (see SM):
\begin{align}
E_{\vec{n}, \vec{n} + \vec{\mu}}  = 2^{-1} + \sum_{i=0}^{N_r-1} 2^{i-1} Z^{(i)}_{\vec{n}, \vec{n} + \vec{\mu}} \, .
\end{align}
Every other operator in the Hamiltonian is constructed using the projectors $\mathcal{P}^{(i)}_{\vec{n}, \vec{n} + \vec{\mu}}$. 
As a result, every rotor is exactly encoded by $N_r$ qubits, 
while ensuring the preservation of gauge symmetry (final panel in Fig~\ref{fig:RishonConstruction}).

\begin{figure*}
    \centering
    \includegraphics[width=\textwidth]{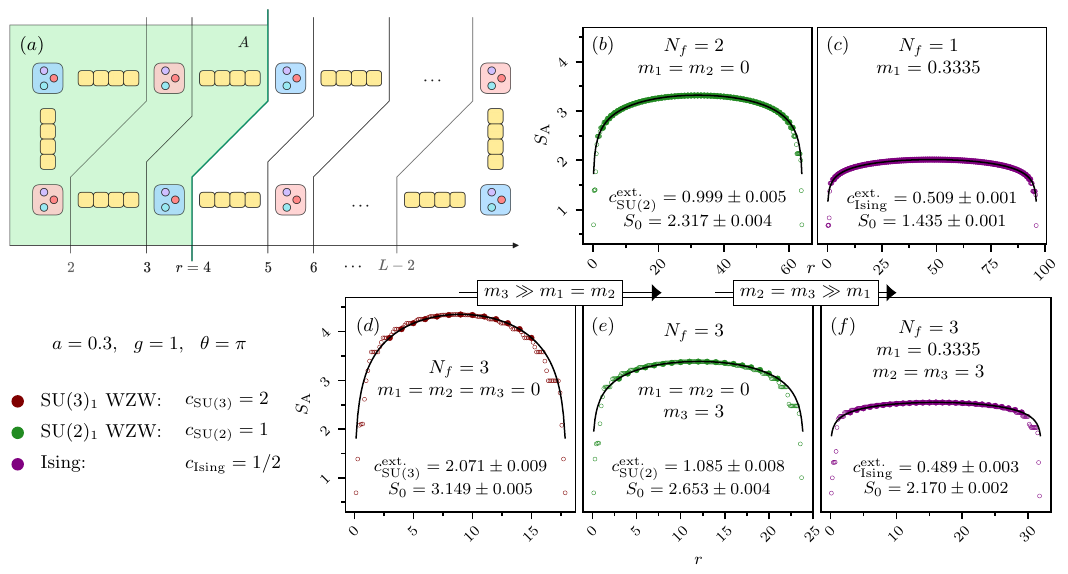}
    \vspace{-2em}
    \caption{\label{fig:CentralCharge} Variation of entanglement entropy with subsystem-size~$r$ for the Schwinger model with $N_f$ fermions on a ring. ($a$) The system consists of a ring of size $L \in \{16,24,32,64,96\}$ unit cells.
    Each cell has $N_f \leq 3$ and $N_r=4$ qubits to encode the rotor. At the quantum-critical point, the central charge is extracted by fitting the numerical data to Eq.~\eqref{eq:CardyFormula}. The expected central charges for the different quantum-critical points described by~${\rm SU}(N_f)_1$ WZW models are shown. Only bipartitions including full unit cells were considered in the fits.
    ($b$, $c$) For $N_f=2$ and $N_f=1$ with $\theta=\pi$ the obtained central charges are close to the expected results $c_{\rm SU(2)}=1$ and the $c_{\rm Ising}=1/2$ respectively. ($d$) For $N_f=3$ and~$\theta=\pi$, the  central charge is again obtained to be close to $c_{\rm SU(3)}=2$. ($e$, $f$) One or two of the masses are made much heavier than the rest to reach the~$SU(2)_1$, and the Ising critical points, as a crosscheck of our results. The relatively strong deviation from $c_{\rm Ising}=1/2$ in panel $(f)$ is due to the $N_f$ dependence of the mass shift $\tilde{m}_\alpha = m_\alpha - ag^2N_f/8$~\cite{Dempsey:2022nys}; more details in the SM. While data are shown only for the renormalization group fixed points, results for the entire flow  can be obtained similarly. Note that the quoted errors are from the fit only, see main text for further discussion.
    }
\end{figure*}

\emph{Gapless points of the multi-flavor Schwinger model.} 
To illustrate the computational efficiency of the VR binary encoding,  we first perform Matrix Product State (MPS) simulations in the $d=1$ theory with multiple fermion flavors. This is the lattice multi-flavor Schwinger model, whose rich phenomenology~\cite{Dempsey2024,steinhardtSU2flavorSchwingerModel1977,Hetrick:1995wq, Banuls:2016gid,Coleman1976} has several features in common with QCD. We will focus on the gapless points of the model at finite $\theta$, generated by shifting the electric field operator in $H$ by $\theta/(2\pi)\leq 1$, due to the presence of a quantum chiral anomaly~\cite{Coleman:1975pw,Coleman1976}.

 For $N_f=1$, the model has a phase transition of the Ising universality class  at the critical mass value $m \approx 0.3335$  for $\theta=\pi$~\cite{byrnesDensityMatrixRenormalization2002,arguellocruzPrecisionStudyMassive2025, fujiiCriticalBehaviorSchwinger2025}. This is characteristic of a confinement--deconfinement transition with emergent half-asymptotic excitations, corresponding to central charge  $c_{\rm Ising}=1/2$~\cite{Coleman1976,Mandelstam:1975hb}. This value is verified  by varying the entanglement entropy with subsystem size~[Fig.~\ref{fig:CentralCharge} ($a$)], using the standard formula~\cite{Calabrese2008}:
 \begin{align}
      S_A=\frac{c}{3} \log \left(\frac{L}{\pi}\sin \frac{\pi r}{L}\right) + S_0\, \label{eq:CardyFormula}\,.
 \end{align}
 Here $r$ denotes the spatial extent of subregion $A$ within a system of size $L$, in units of staggered cells. The central charge is denoted by~$c$ and~$S_0$ contains non-universal lattice dependent contributions. The numerical result  shown in Fig.~\ref{fig:CentralCharge}(c) was obtained by constructing the ground-state with a four-qubit encoding of the model using the density matrix renormalization group (DMRG) algorithm implemented in TeNPy~\cite{hauschildEfficientNumericalSimulations2018,hauschildTensorNetworkPython2024}.
 Both open and periodic boundary conditions are readily implemented in the framework using a folded MPS ordering. The latter eliminates boundary effects, allowing one to extract the central charge without elaborate finite-size scaling analyses~\cite{Dempsey2024, campostriniFinitesizeScalingQuantum2014}; a detailed study is provided in the SM. 
 We extract $c^{\rm ext.}_{\rm Ising}=0.509\pm0.001$ from a single wavefunction where the quoted error is only from the fit; see~\cite{byrnesDensityMatrixRenormalization2002,arguellocruzPrecisionStudyMassive2025, fujiiCriticalBehaviorSchwinger2025,Dempsey2024} for other recent numerical results using tensor network methods.  Even though we eliminate spurious boundary effects --- Fig. S3 in SM compares the corresponding data using open boundary conditions --- systematic finite size, MPS bond dimensions or lattice artifact effects remain. The latter lead to deviations from the theoretical predictions. Some of these can be quantified following Refs.~\cite{Cardy:2010zs, Roy:2025hew}.

Extending our framework to $N_f>1$, the massless Schwinger model exhibits a split between an anomaly--lifted axial singlet and an anomaly--free non--singlet sector, mirroring the separation between the $\eta'$ and the pion/$\eta$ channels in QCD~\cite{Veneziano:1979ec,Adler:1969gk,Bell:1969ts,tHooft:1976rip,tHooft:1976snw,Witten:1979vv}. 
In the continuum, and at low energies, the theory factorizes into an $\mathrm{SU}(N_f)_1$ Wess--Zumino--Witten (WZW) theory at level $k=1$~\cite{gepnerNonabelianBosonizationMultiflavor1985,affleckRealizationChiralSymmetry1986}, describing the massless non--singlet sector (pion/$\eta$ modes), and a remaining $\mathrm{U(1)}_A$ sector with the anomaly--induced massive Schwinger boson ($\eta'$ mode). The WZW sector is characterized by a central charge $c_{{\rm SU}(N_f)}=N_f-1$. In appropriate massive decoupling limits, the theory flows to the $N_f=1$ Ising critical point. The numerical extraction of WZW charges for $N_f>1$ is substantially more demanding, making it an ideal benchmark to illustrate  the computational power of the VR construction.

In Fig.~\ref{fig:CentralCharge} (d)-(f), we show results for the relevant central charges for $N_f=3$.
For purely massless fermions, we obtain good agreement with the expected value for the central charge: $c^{\rm ext.}_{\rm SU(3)}=2.071\pm0.009$. As discussed in the SM, this is the most computationally intensive simulation. Taking $m_3\gg m_{1,2}$, we extract the WZW central charge for $N_f=2$ to be $c^{\rm ext.}_{\rm SU(2)}=1.085\pm0.008$. 
A direct extraction in the $N_f=2$ model [panel (b)] gives $c^{\rm ext.}_{\rm SU(2)}=0.999\pm0.005$, demonstrating the robustness of the results. Finally, taking the massive limit for $m_{3,2}$ and setting $m_1$ to the critical value in the $N_f=1$ model, we find $c^{\rm ext.}_{\rm Ising}=0.489\pm0.003$ in agreement with the direct extraction shown in panel (c); see~\cite{Dempsey2024} for a recent  $N_f=2$ computation and also earlier for discussion of the discrepancies to theoretical values. 

In these simulations, we employed  $N_\mathrm{qubits} = (N_f + N_r)L \propto \mathcal{O}(100)$. Only physical cuts between unit cells were considered for the fit, with the first and last 2 unit cells excluded to minimize finite size effects. 
To account for this, as well as errors introduced due to tensor network truncation, system sizes were chosen to be as big as possible while keeping truncation errors below $2.5\times10^{-6}$. We used maximal bond dimensions between $\chi=8192$ for $N_f = 3$ and $\chi=2048$ for $N_f=1$, see the SM. 

\emph{String tension in $d=2$ pure gauge theory.}
In $d=1$, dynamical simulation of the gauge field can be avoided by introducing nonlocal all-to-all couplings~\cite{Hamer:1997dx} for open boundary conditions. This is not possible in higher spatial dimensions, allowing for a further test of the VR formalism. To demonstrate the robustness of our method, we will compute the string tension of confined electric fluxes in U(1) gauge theory without matter fields.

\begin{figure}
    \centering
    \vspace{-1em}
    \includegraphics{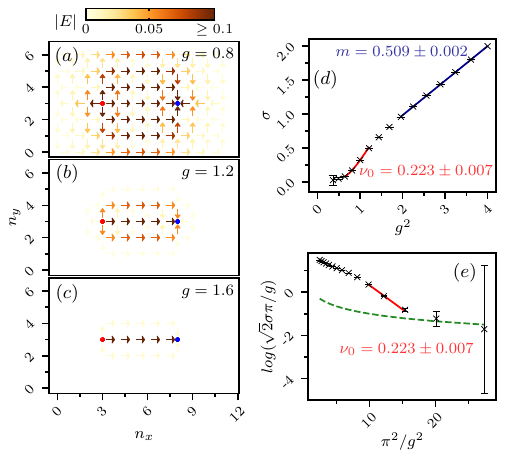}
    \vspace{-3em}
    \caption{String tension in $d=2$ U(1) gauge theory without matter fields. $(a)$--$(c)$: 
Red and blue dots with separation $\Delta=5$ represent background charges. 
With increasing coupling constant $g$, 
 the electric field profile narrows, resulting in a flux tube.  $(d)$: Fit to linear scaling of  ground-state energy with $\Delta$, allowing extraction of $\sigma$ for large coupling. The expected $g^2/2$ scaling of $\sigma$~\cite{Kogut:1974ag} is verified (blue fit). Gaussian decay is expected for small $g$~\cite{hamerWeakcouplingExpansionsEffective1993} (red fit). $(e)$:  Best extracted value for decay constant is  $\nu_0=0.223$ (red fit), to be compared to expected $\nu_0=0.321$~\cite{loanPathIntegralMonte2003}. Finite size effects are estimated by $\sigma_\mathrm{min} = g^2/(2L_y)$~\cite{hamerWeakcouplingExpansionsEffective1993} (green dashed line).
   } 
    \label{fig:QED3}
\end{figure}

Flux string configurations are generated in the presence of two opposite static background charges on the lattice, and are characterized by a coupling-dependent string tension $\sigma$. While at strong coupling, the electric field is localized along the classical path, its profile expands with decreasing coupling constant $g$. This feature is cleanly exhibited in Fig.~\ref{fig:QED3}(a)-(c). To quantify these features, we consider the energy added by the electric flux tube to the ground-state energy, expected to be proportional to the distance $\Delta$ between the background charges:
\begin{align}
    E_0(\Delta) = \sigma\Delta + E_0 \, .\label{eq:QED3_energy_String_tension}
\end{align}
The model experiences a crossover from strong to weak coupling, with the 
former given by $\sigma= g^2/2$ \cite{Kogut:1974ag}, and the latter by $\sigma \propto \sqrt{g^2/\pi^2}\exp(-\nu_0 \pi^2/g^2)$, with $\nu_0 \approx 0.321$ \cite{loanPathIntegralMonte2003}.
The weak coupling expression reflects monopole-instanton ~\cite{Polyakov:1975rs,Polyakov:1976fu} confining dynamics for $d=2$, which generates the nonperturbatively small string tension.  

In contrast to $d=1$, rapidly growing entanglement between a spatial region and the rest of the system in two spatial dimensions leads to exponential growth in required computational resources. Conservation of local U(1) charges in the VR formalism reduces computational demands significantly. Specifically,  simulations of grid sizes up to $12\times7$ and rotor dimension of 16, encoded using 4 qubits, with a total $\mathcal{O}(300)$ qubits, and total bond-dimension of up to $\chi=8192$ are feasible, using only up to 32 GB RAM.  This reduces computational times to the order of days, rather than weeks, using maximally 24 cores (see SM for more details).

Using DMRG for different $\Delta$, the string tension is extracted using Eq.~\ref{eq:QED3_energy_String_tension}, and is shown in Fig. \ref{fig:QED3}(b). At large couplings, the behavior matches the expected $g^2/2$ value closely; in contrast, the results deviate significantly from the small coupling limit, with the best result for $g \in [0.8, 1.0]$ being $\nu_0 = 0.223$ to be compared to the expected value of $\nu_0=0.321$ \cite{loanPathIntegralMonte2003}; see also Ref.~\cite{benderRealtimeDynamics$2+1D$2020} for a similar discussion. The weak coupling regime is especially challenging as the exponentially decaying $\sigma$ requires increasingly higher accuracy. Furthermore, increased delocalization introduces boundary effects and larger maximal rotor dimension; entanglement increases substantially too, leading to higher truncation errors. This is reflected in the error bars of Fig \ref{fig:QED3}(a). These limitations are of the classical tensor network simulation method and not inherent to the VR formalism. Further improvements can therefore be achieved from larger scale classical computations or implementation on quantum hardware -- see Ref.~\cite{ciavarella2025} for quantification of digitization/truncation errors in terms of the electric basis.

\emph{Conclusion.} 
  We introduced a novel virtual rishon framework for classical and quantum simulation of quantum field theories with gauge constraints. A key feature is the exact conservation of gauge symmetries via an intermediate quantum--link representation, which also enables qubit encoding of  gauge and matter degrees of freedom. The framework reduces the computational complexity of the simulations by factorizing into sectors for an extensive number of conserved quantities and by avoiding mixing matter and gauge fields into combined sites, keeping the local Hilbert spaces separate. 

We illustrated the robustness of this VR framework for lattice gauge theories in $d=1$ space dimension by systematically extracting the central charges of the critical points of the $N_f\leq3$ flavor Schwinger model. For $d=2$, we computed the string tension of electric flux tubes connecting static charges. All our computations were performed using qubit-encoded rotors and thus, can be implemented on quantum hardware using the algorithms developed in Refs.~\cite{Roy2024ec, Roy2023efficient, Roy:2024xdi, Rogerson2024, Lamb:2025rbl}. Finally, the VR formalism can be  extended to non-abelian gauge theories, preserving the unique features of the construction. We leave a detailed analysis of this important generalization to a future work.

\emph{Acknowledgments.}
The authors thank Sidan A, Andreas Gleis, Pradeep Kattel and Thomas Banks for useful discussions. AR, DR, and RMK were supported by the U.S. Department of Energy, Office of Basic Energy Sciences, under Contract No. DE-SC0012704.
R.V. is supported by the U.S. Department of Energy, Office of Science under contract DE-SC0012704. His work on quantum information science is supported by the U.S. Department of Energy, Office of Science, National Quantum Information Science Research Centers, Co-design Center for Quantum Advantage (C$^2$QA) under contract number  DE-SC0012704. R.V. was also supported at Stony Brook by the Simons Foundation as a co-PI under Award number 994318 (Simons Collaboration on Confinement and QCD Strings). R.V. acknowledges support from the Royal Society Wolfson Foundation Visiting Fellowship and the hospitality of the Higgs Center at the University of Edinburgh. 

\bibliography{FinalRefs}

\clearpage
\onecolumngrid

\appendix
\section{\huge{Supplemental Material}}

\setcounter{secnumdepth}{3}

\setcounter{equation}{0}
\setcounter{figure}{0}
\setcounter{table}{0}
\setcounter{page}{1}

\makeatletter
\renewcommand{\theequation}{S\arabic{equation}}
\renewcommand{\thefigure}{S\arabic{figure}}
\renewcommand{\thetable}{S\arabic{table}}
\renewcommand{\@appendixcntformat}{Supplement \thesection:\ } 
\makeatother

\section{Overview}
This supplemental material is organized into four parts. 
\begin{enumerate}
    \item In Appendix B we provide a more detailed explanation of the virtual rishon formalism. This includes details on the implementation of gauge charge conservation and the derivation of the necessary projection operators. We conclude this section with a discussion of the qubit encoding. 
    \item Appendix C expands on the numerical results provided for the massive Schwinger model. We provide additional motivation for the explicit conservation of gauge charges in this model, a discussion on boundary effects, as well as a discussion on the $N_f$ dependence of the chiral symmetry restoring mass shift \cite{Dempsey:2022nys}. We conclude the appendix with a convergence analysis of the DMRG data presented in the main text. 
    \item Appendix D focuses on the results of the QED$_3$ data. We show the extrapolation of the ground-state energy towards infinite MPS bond dimension and the extraction of the string tension $\sigma$ used in the main text.
    \item The final Appendix E provides information on the source code and data availability of this publication.
\end{enumerate}

\section{Details on the Virtual Rishon formalism}
We outlined the virtual rishon construction in Fig.~\ref{fig:RishonConstruction} of the main text. The relevant steps are described in more detail below. 

\subsection{Details of Step $2)$  in Fig.~\ref{fig:RishonConstruction} : Identifying Non-overlapping Local U(1) Charges}

\label{sec:ChargeConservationWithTensorNetworks}
Tensor network libraries~\cite{hauschildEfficientNumericalSimulations2018, hauschildTensorNetworkPython2024, fishmanITensorSoftwareLibrary2022} form the backbone of algorithms like DMRG~\cite{White1992, Schollwock2011}. These libraries allow the parametrization of tensors while conserving abelian~\cite{singhTensorNetworkDecompositions2010, singhTensorNetworkStates2011} and non-abelian \cite{devosTensorKitjlJuliaPackage2025, weichselbaumNonabelianSymmetriesTensor2012} charges. This is achieved by identifying the symmetry properties of the different tensor indices and algorithmically deducing the allowed irreducible representations of the corresponding tensor. These are then represented block diagonally. In classical tensor network calculations, this not only reduces the memory footprint but also reduces the computational complexity; in hybrid classical-quantum algorithms, the reduction of parameters can yield favorable optimization results \cite{Roy2024ec, Roy2023efficient, Roy:2024xdi, Rogerson2024, Lamb:2025rbl}. In this work, we use this proven computational method to enforce gauge charge conservation.
We focus on the implementation where tensors are allowed to carry a charge themselves; this is the convention in TeNPy~\cite{hauschildTensorNetworkPython2024} and differs from other implementations, e.g. Tensorkit.jl~\cite{devosTensorKitjlJuliaPackage2025}. The two approaches are equivalent for abelian charges if the extra charge is carried by an extra virtual index of dimension $1$.
In the following, we will review a simple charge counting rule for the Schwinger model which is sufficient for identifying the abelian representations discussed here.  It consists of four steps:
\begin{enumerate}
    \item Choose the computational basis in which the conserved charges are diagonal
    \item Identify how each single site operator used in the terms of the Hamiltonian connects different charge values.
    \item Identify $\delta Q$, the amount by which the charges change.
    \item Save the block matrix form for the single-site operator with $\delta Q$.
    \item Define the Hamiltonian in terms of these block matrices; the total charge $\sum_i \delta Q_i = 0$ of each multi-site operator $h = \bigotimes_i O_i$ must vanish.
\end{enumerate}

In the Schwinger model example, we have one U(1) charge per unit cell, which we label as the vector
\begin{align}
    \vec{Q} = \{ G_1 \dots G_L \}\,,
\end{align}
where the $\vec{n}$ and $\vec{\mu}$ (see Eq.~\eqref{eq:gaugeOpRishon})is dropped for the 1D case:
\begin{align}
    G_{n} =  \sum_\alpha^{N_{f}} c^{(\alpha) \dagger}_{n}c^{(\alpha)}_{n} +\left(N_{a,n} +N_{b,n} - \bar{N}\right)\, .
\end{align}
This suggests that the Fock basis serves as the computational basis for both fermions and bosonic rishons. Consider occupation number operators:
For fermions $c^{(\alpha) \dagger}_{n}c^{(\alpha)}_{n}$, the local operator connects the $n$-th charge in a diagonal fashion
\begin{align}
     G_n:&\  0 \rightarrow 0 \text{ with matrix element } 0\,,\\
     G_n:&\  1 \rightarrow 1 \text{ with matrix element } 1\,.
\end{align}
It acts trivially on all other charges.
Similarly for the bosonic rishon occupation $N_{a/b,n}$
\begin{align}
     G_n:&\  0 \rightarrow 0 \text{ with matrix element } 0\,,\\
     G_n:&\  1 \rightarrow 1 \text{ with matrix element } 1\,,\\
     &\vdots \notag \\
     G_n:&\  N_{max} \rightarrow N_{max} \text{ with matrix element } N_{max}\,.
\end{align}
The ladder operators $c^{(\alpha) \dagger}_n, a^\dagger_n$ and $b^\dagger_n$, on the other hand, all increase the $n$-th component of the charge vector by $\delta Q_n = 1$ while acting trivially on all other components.
For the fermionic operators $c^{(\alpha) \dagger}_n$, this leads to
\begin{align}
     G_n:&\  0 \rightarrow 1\text{ with matrix element } 1\,,\\
\end{align}
as the only matrix elements, and for the bosonic rishon rotors $a^\dagger_n$ and $b^\dagger_n$ in 
\begin{align}
     G_n:&\  0 \rightarrow 1 \text{ with matrix element } 1\,,\\
     G_n:&\  1 \rightarrow 2 \text{ with matrix element } 1\,,\\
     &\vdots \notag \\
     G_n:&\  N_{max}-1 \rightarrow N_{max} \text{ with matrix element } 1 \, .
\end{align}
The annihilation operators behave analogously, but with the opposite effect on the charges.
After the charge mappings are identified, the irreducible representations of the occupation-number operators are just lists of one-dimensional representations of their local charge.
For example, for the bosonic rishons,~$N_{a/b, n}$ is given by $N_r$ blocks:
\begin{align}
    N_{a/b, n}= \begin{bmatrix}
        0 & & & \\
        &1 & & \\
        & & \ddots &\\
        & & & N_{max}
    \end{bmatrix}.
\end{align}
The representation of multi-site operators can be derived automatically from the single-site operators as long as the constraint $\sum_i \delta Q_i = 0$ is satisfied.  This restricts the different multi-site operators that can be constructed. For instance, a bosonic creation operator $a^{\dagger}_{n}$ must be paired to either a fermionic $c_{n}^{(\alpha)}$ or a rishon annihilation operator $b_n$ acting on the same unit cell. This automatically ensures that only terms similar to the ones used in Eq.~\eqref{eq:Hamiltonian} are allowed. While terms like $c_{n}^{(\alpha)}a^{\dagger}_n$ could exist based on this gauge symmetry, the quantum link model representation is only valid under the conservation of another local constraint, the link constraint $N_a + N_b=\bar{N}$, which forbids these two site interactions. One of the main ideas of this work is to use projectors to enforce this constraint, as described in the next section.

\subsection{\label{sec:ProjectorLinkConstraint} Details of  Step $3a)$ in in Fig.~\ref{fig:RishonConstruction}: Projection of the Link Constraints}
At the core of the virtual rishon formulation is the idea that the quantum link model rishons are only an intermediate tool and that the rishon pairs are projected back to a single rotor degree of freedom once the Hamiltonian is parametrized in terms of the irreducible representations. 
The correct projection operators that enforce the local link constraints $\bar{N} = N^{(\mu)}_{a,\vec{n}} + N^{(\mu)}_{b, \vec{n} + \vec{\mu}}$ can be identified by representing the constraint as a set of eigenvalue equations
\begin{align}
 \left( N^{(\mu)}_{a,\vec{n}} + N^{(\mu)}_{b,\vec{n} + \vec{\mu}} \right)\ket{N}_{a, \vec{n}}^{(\mu)} \ket{N}^{(\mu)}_{b, \vec{n} + \vec{\mu}} =\bar{N} \ket{N}_{a, \vec{n}}^{(\mu)} \ket{N}^{(\mu)}_{b, \vec{n} + \vec{\mu}}\,.
\end{align}
Here $\ket{N}^{(\mu)}_{a/b, \vec{n}}$ label the rishon states in the occupation basis.
Each of these local equations is solved in the $\bar{N}+1$ dimensional local subspace
\begin{align}
\left\{ \ket{N_a}_{a, \vec{n}}^{(\mu)} \ket{\bar{N}-N_a}^{(\mu)}_{b, \vec{n} + \vec{\mu}}  \vert N_a \in [0,1, \dots , \bar{N}] \right\}\,.
\end{align}
Furthermore, an orthonormal basis of that subspace can be found that is diagonal with respect to the original gauge field occupation
\begin{align}
    E_{\vec{n},\vec{n}+\vec{\mu}}\ket{N_a}_{a, \vec{n}}^{(\mu)} \ket{\bar{N}-N_a}^{(\mu)}_{b, \vec{n} + \vec{\mu}} = \frac{1}{2}\left( \bar{N} - 2N_a\right)\ket{N_a}_{a, \vec{n}}^{(\mu)} \ket{\bar{N}-N_a}^{(\mu)}_{b, \vec{n} + \vec{\mu}} = e\ket{N_a}_{a, \vec{n}}^{(\mu)} \ket{\bar{N}-N_a}^{(\mu)}_{b, \vec{n} + \vec{\mu}} \, .
\end{align}
These eigenvalues are chosen as state labels of the projected reduced basis. 
While for odd $\bar{N}$ the electric field operator takes half-integer eigenvalues, we assume here even $\bar{N}$ to ensure the usual integer representation.
Finally, a change of variables $e = \frac{\bar{N}}{2} - N_a$, defines the projection operator onto the space with the enforced link constraint used in the main text:
\begin{align}
P_{\vec{n}, \vec{n} + \vec{\mu}} = \sum_{e = -\bar{N}/2}^{\bar{N}/2} \ket{e}_{\vec{n}, \vec{n} + \vec{\mu}}\bra{\bar{N}/2-e}_{a, \vec{n}}^{(\mu)}\bra{\bar{N}/2+e}^{(\mu)}_{b, \vec{n} + \vec{\mu}} \, . 
\end{align}
The process of projecting the Hamiltonian back to the rotor basis is described in the main text. For our DMRG results, we project each single-site operator used in the Hamiltonian individually onto the reduced basis and construct the Hamiltonian directly in the projected representation. The labels $\{ \ket{e} \}$ represent the reduced basis of the gauge rotor Hilbert space, and the projected operators: 
\begin{align}
    \tilde{E}_{\vec{n},\vec{n}+\vec{\mu}} &= \frac{1}{2}P_{\vec{n},\vec{n}+\vec{\mu}}\left(N_{b, \vec{n}+ \vec{\mu}} -N_{a, \vec{n}} \right) P^\dagger_{\vec{n},\vec{n}+\vec{\mu}}\\
    \tilde{U}_{\vec{n},\vec{n}+\vec{\mu}} &=  P_{\vec{n},\vec{n}+\vec{\mu}} a^{(\mu)}_{\vec{n}}b^{(\mu)\dagger}_{\vec{n}+\vec{\mu}} P_{\vec{n},\vec{n} + \vec{\mu}}^{\dagger}
\end{align} form the algebra on that Hilbert space.
The rishon formulation is therefore equivalently a construction of the gauge-constrained Hilbert space, rather than just the specific Hamiltonian.

\subsection{Details of the Qubit Encoding Step $1^*)$ and $3b)$ in Fig.~\ref{fig:RishonConstruction}}
To have the virtual rishon formulation be amenable to implementation on existing quantum computing hardware, we performed the simulations using a binary encoding of the rotor degree of freedom - as shown in Fig.~\ref{fig:RishonConstruction} of the main text.
This binary encoding into $N_r$ qubits can be performed individually on each rishon  as shown in step 1*:
\begin{align}
    N_{a/b, \vec{n}}^{(\mu)} = \sum_{i=0}^{N_r-1} 2^{i} n_{a/b, \vec{n}}^{(i,\mu)}
\end{align}
\begin{figure}
        \includegraphics{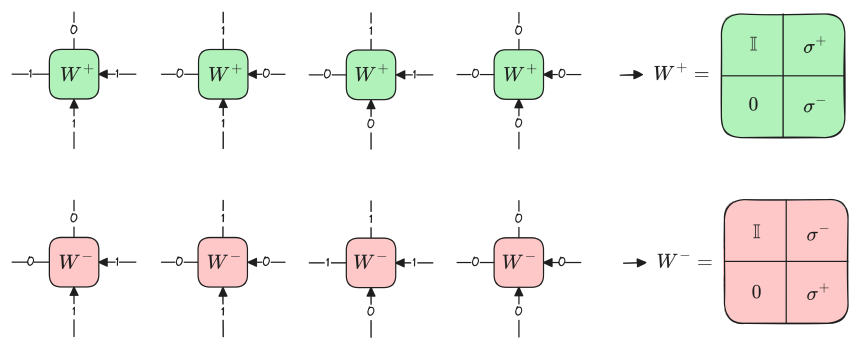}
    \caption{\label{fig:BinaryEncodedLadder} Ladder operator of the qubit encoded rotor}
\end{figure}

The operators $a^{(\dagger)}$ and $b^{(\dagger)}$~(position labels suppressed for the ease of presentation), which are now $N_r$-qubit operators are more involved. While there are various ways to derive the necessary multi-site operator, here we use a matrix product operator (MPO) representation. 
The adder is composed of a product of two qubit operators, one qubit represents the current physical binary power qubit, while the other is the incoming and outgoing ``carry" on the qubit. 
The operation on the qubit pair is simple if the incoming carry qubit is $\ket{0}$ the operator acts as an identity, if it is $\ket{1}$ then it should map $\ket{0}_p\ket{1}_c\rightarrow \ket{1}_p\ket{0}_c$ and $\ket{1}_p\ket{1}_c\rightarrow \ket{0}_p\ket{1}_c$, where the outgoing carry bit represents the binary overflow to the next qubit power.
Concatenating the outgoing carry qubits to the incoming one of the next binary power reveals that the carry qubit is effectively a virtual index of an MPO representation, see Fig. \ref{fig:BinaryEncodedLadder} for a graphical overview.

The fundamental units of the MPO are matrices of Pauli operators
\begin{align}
    &W^+ = \begin{bmatrix}
    \mathbb{I}& \sigma^+\\
    0 & \sigma^-
    \end{bmatrix} \,.
\end{align}
The operator reducing the occupation is derived similarly
\begin{align}
    W^- = \begin{bmatrix}
    \mathbb{I}& \sigma^-\\
    0 & \sigma^+
    \end{bmatrix}\,,
\end{align}
such that the full $N_r$ qubit ladder operators takes the form 
\begin{align}
    a^\dagger &= \begin{bmatrix}
    1 & 0\\
    \end{bmatrix} \left( W^{+} \right) ^{N_r} \begin{bmatrix}
    0 \\ 1
    \end{bmatrix}\,,\\
a &= \begin{bmatrix}
    1 & 0\\
    \end{bmatrix} \left( W^{-} \right) ^{N_r} \begin{bmatrix}
    0 \\ 1
    \end{bmatrix}\,.
\end{align}
The right boundary vector indicates that we want to add/subtract exactly one, while the left ensures that over/undercounting results in a $ 0$-valued total operator.
Expanding the operator either by matrix multiplication or by inspection of the MPO state machine representation \cite{PhysRevA.78.012356}, results in a sum with $N_r$ terms, spanning up to $N_r$ qubits:
\begin{align}
    a^\dagger = \sum_{i=0}^{N_r-1} \mathbb{I}^{\otimes N_r-i-1}  \sigma^{+}\left(\sigma^{-}\right)^{\otimes i}\label{eq:BinaryLadder}
\end{align}
Similarly for $a, b$ and $b^\dagger$.
Continuing to step 2, gauge symmetry conservation requires only minimal modification to the non-qubit encoded case; the individual qubits contribute with respect to their respective binary power to the total U$(1)$ gauge charge:
\begin{align}
    G_{\vec{n}} =  \sum_\alpha^{N_{f}} c^{(\alpha) \dagger}_{\vec{n}}c^{(\alpha)}_{\vec{n}} + \sum_{\vec{\mu}}\left[ \sum_{i=0}^{N_r-1}\left[2^{i}\left(n^{(i,\mu)}_{a,\vec{n}} +n^{(i,\mu)}_{b,\vec{n}} \right)\right] - \bar{N} \right]\, \label{eq:gaugeOpRishonQubit}
\end{align}

A much larger modification is made to the projector used in step 3b. Assuming $N_r$ qubits, the maximum possible occupation per rishon is $2^{N_r}-1$. In the quantum link model formulation, the effective rotor dimension is governed by the space spanned via the link constraint: $\bar{N} = N_a + N_b $. While other choices $\bar{N} \leq 2^{N_r}-1$ are possible, setting $\bar{N} = 2^{N_r}-1$ results in the maximal encoding density.
Remarkably, this case results also in a major simplification of the Projector $P^{\bar{N}}$ that enforces the link constraint.
Representing the constraint again (see SM Sec. \ref{sec:ProjectorLinkConstraint}) in terms of an operator equation
\begin{align}
 \left( N^{(\mu)}_{a,\vec{n}} + N^{(\mu)}_{b,\vec{n} + \vec{\mu}} \right)\ket{N}_{a, \vec{n}}^{(\mu)} \ket{N}^{(\mu)}_{b, \vec{n} + \vec{\mu}} =\bar{N} \ket{N}_{a, \vec{n}}^{(\mu)} \ket{N}^{(\mu)}_{b, \vec{n} + \vec{\mu}}\,,
\end{align}
as $2^{N_r}-1$ in binary encoding is represented  by  $N_r$ ones, for example, $111\dots11$, the binary encoding results in
\begin{align}
 \left[\sum_{i=0}^{N_r-1}\ 2^i\left( n^{(i,\mu)}_{a,\vec{n}} + n^{(i,\mu)}_{b,\vec{n} + \vec{\mu}} \right)\right]\bigotimes_{i=0}^{N_r-1}\ket{n}_{a, \vec{n}}^{(i,\mu)} \ket{n}^{(i,\mu)}_{b, \vec{n} + \vec{\mu}} =\left[\sum_{i=0}^{N_r-1} 2^i \cdot 1 \right]\bigotimes_{i=0}^{N_r-1}\ket{n}_{a, \vec{n}}^{(i,\mu)} \ket{n}^{(i,\mu)}_{b, \vec{n} + \vec{\mu}}\,. \label{eq:BinaryProjector}
\end{align}
Generally, $n^{(i)}_a + n^{(i)}_b$ can take the values [0,1,2],  where 2 would create overflow into the next binary power. In this case, however, it can be shown that $n^{(i)}_a + n^{(i)}_b = 1$ for all binary powers $i$. Starting with $i=0$ the only way to satisfy the left hand side of \eqref{eq:BinaryProjector} is
\begin{align}
    2^0\left( n^{(i,\mu)}_{a,\vec{n}} + n^{(i,\mu)}_{b,\vec{n} + \vec{\mu}} \right) \ket{n^{(0)}}_{a, \vec{n}}^{(\mu)} \ket{n^{(0)}}^{(\mu)}_{b, \vec{n} + \vec{\mu}} =2^0 \ket{n}_{a, \vec{n}}^{(0,\mu)} \ket{n}^{(0,\mu)}_{b, \vec{n} + \vec{\mu}}\,,\\
    \left( n^{(i,\mu)}_{a,\vec{n}} + n^{(i,\mu)}_{b,\vec{n} + \vec{\mu}} \right) \ket{n^{(0)}}_{a, \vec{n}}^{(\mu)} \ket{n^{(0)}}^{(\mu)}_{b, \vec{n} + \vec{\mu}} =1 \ket{n}_{a, \vec{n}}^{(0,\mu)} \ket{n}^{(0,\mu)}_{b, \vec{n} + \vec{\mu}}\,.
\end{align}
The l.h.s in parenthesis=1 ensures that the equation is self-contained and does not affect the equations of subsequent binary powers. Repeating this observation for the subsequent powers results in a perfect factorization of the total link constrained subspace, resulting in:
\begin{align}
 \left[\bigotimes_{i=0}^{N_r-1}\ 2^i\left( n^{(i,\mu)}_{a,\vec{n}} + n^{(i,\mu)}_{b,\vec{n} + \vec{\mu}} \right)\ket{n}_{a, \vec{n}}^{(i,\mu)} \ket{n}^{(i,\mu)}_{b, \vec{n} + \vec{\mu}}\right] =\left[\bigotimes_{i=0}^{N_r-1} 2^i \cdot 1 \ket{n}_{a, \vec{n}}^{(i,\mu)} \ket{n}^{(i,\mu)}_{b, \vec{n} + \vec{\mu}}\right]\,.
\end{align}
Each of these equations is solved via the two-dimensional subspace of odd qubit parity
\begin{align}
    \left\{  \ket{0}_{a, \vec{n}}^{(i,\mu)} \ket{1}^{(i,\mu)}_{b, \vec{n} + \vec{\mu}},\, \ket{1}_{a, \vec{n}}^{(i,\mu)} \ket{0}^{(i,\mu)}_{b, \vec{n} + \vec{\mu}} \right\}\,.
\end{align}
Following SM Sec. \ref{sec:ProjectorLinkConstraint}, the electric field operator is diagonal with respect to these two states, which either contribute positively or negatively with their respective binary power to the total electric field. Using $Z_{\vec{n} + \vec{\mu}}^{(i, \mu)}$ as the local Pauli-Z of the corresponding binary power $i$ we choose the labels
\begin{align}
    \ket{\uparrow}_{\vec{n} + \vec{\mu}}^{(i, \mu)} = \ket{0}_{a, \vec{n}}^{(i,\mu)} \ket{1}^{(i,\mu)}_{b, \vec{n} + \vec{\mu}}\,,\\
    \ket{\downarrow}_{\vec{n} + \vec{\mu}}^{(i, \mu)} = \ket{1}_{a, \vec{n}}^{(i,\mu)} \ket{0}^{(i,\mu)}_{b, \vec{n} + \vec{\mu}}\,.
\end{align}
With this, the projector on to the total link-constrained space 
\begin{align}
    P^{\bar{N}}_{\vec{n} + \vec{\mu}} = \bigotimes_{i=0}^{N_r-1} \mathcal{P}^{(i)}_{\vec{n} + \vec{\mu}}\,,
\end{align}
where each qubit power is individually projected via
\begin{align}
\mathcal{P}^{(i)}_{\vec{n}, \vec{n} + \vec{\mu}} &= \ket{\uparrow}^{(i)}_{\vec{n}, \vec{n} + \vec{\mu}}   \bra{0}^{(i,\mu)}_{a,\vec{n}} \bra{1}^{(i,\mu)}_{b,\vec{n} + \vec{\mu}} \nn &+ \ket{\downarrow}^{(i)}_{\vec{n}, \vec{n} + \vec{\mu}}   \bra{1}^{(i,\mu)}_{a,\vec{n}} \bra{0}^{(i,\mu)}_{b,\vec{n} + \vec{\mu}}\; . \label{eq:ProjQubit}
\end{align}
One additional subtlety of the qubit encoding is the fact that the electric field is given by 
\begin{align}
 E_{\vec{n}, \vec{n} + \vec{\mu}} = \frac{1}{2}P^{\bar{N}}\left(N^{(\mu)}_{b,\vec{n} + \vec{\mu}} - N^{(\mu)}_{a,\vec{n}}\right) P^{\bar{N}, \dagger}   
\end{align}
which for odd $\bar{N}=2^{N_r}-1$ would result in half-integer electric field values. This can be fixed by introducing a global electric field offset of $1/2$:

\begin{align}
E_{\vec{n}, \vec{n} + \vec{\mu}}  = 2^{-1} + \sum_{i=0}^{N_r-1} 2^{i-1} Z^{(i)}_{\vec{n}, \vec{n} + \vec{\mu}} \, .
\end{align}

For the example of $N_r=3$, this results in the following state labels
\begin{align}
    &E\ket{\uparrow\uparrow\uparrow} = 4\ket{\uparrow\uparrow\uparrow}
    &&E\ket{\uparrow\uparrow\downarrow} = 3\ket{\uparrow\uparrow\downarrow}
    &&E\ket{\uparrow\downarrow\uparrow} = 2\ket{\uparrow\downarrow\uparrow}
    &&E\ket{\uparrow\downarrow\downarrow} = 1\ket{\uparrow\downarrow\downarrow}\\
    &E\ket{\downarrow\uparrow\uparrow} = 0\ket{\downarrow\uparrow\uparrow}
    &&E\ket{\downarrow\uparrow\downarrow} = -1\ket{\downarrow\uparrow\downarrow}
    &&E\ket{\downarrow\downarrow\uparrow} = -2\ket{\downarrow\downarrow\uparrow}
    &&E\ket{\downarrow\downarrow\downarrow} = -3\ket{\downarrow\downarrow\downarrow}\,.
\end{align}
This example reveals that, while not immediate from the definition in Eq. \eqref{eq:BinaryLadder}, the ladder operator  $U_{\vec{n}, \vec{n} + \vec{\mu}} = P^{\bar{N}}a^{(\mu)}_{\vec{n} }b^{(\mu)\dagger}_{\vec{n}+ \vec{\mu}} P^{\bar{N}\dagger}$ and its conjugate, take simple forms in terms in the projected qubit basis as well.
The full projection of all operators is represented in step 3b of Fig~\ref{fig:RishonConstruction}.
Last, but not least, a full qubit encoding also requires the representation of fermions in terms of spins. This is performed using the usual multi-flavor Jordan-Wigner transformation, which is done automatically by the TeNPy library.

\section{Details on the results of the Schwinger model}
The Schwinger model is the simplest nontrivial example of a U(1) gauge theory. Due to its conceptual simplicity, it is studied extensively, and its 1+1D nature allows mappings to nonlocal, gauge-field-independent models, simplifying many computations. Expanding on the results of the main text, we want to show here further subtleties, related to gauge sectors, boundary conditions, and mass shifts.

\subsection{Gauge conservation in the variational ground-state optimization of the Schwinger model}\label{sec:GaugeConservationIsNecessary}
The search for the variational ground-state of the Schwinger model is subject to a subtlety that is easily overlooked in the long-range (LR) formulation. This LR form of the open-boundary Hamiltonian is derived by eliminating the dynamical gauge field using Gauss's Law~\cite{banulsMassSpectrumSchwinger2013}, in particular the identity 
\begin{align}
    G_n\ket{\psi} = q_n\ket{\psi} = N_f\frac{(1-(-1)^n)}{2}\ket{\psi}\,.
\end{align}
Here $G_n$ is the 1+1D version of Eq.\eqref{eq:gausslaw}.
This choice of the background charges $q_n$ is not only the ground-state sector of the weak coupling limit of the massive Schwinger model with $\theta=0$ \cite{banksStrongcouplingCalculationsLattice1976}, but also has been shown to have superior continuum convergence in the presence of a mass-shift \cite{Dempsey:2022nys} constructed to recover discrete chiral symmetry. We emphasize that this choice of $q_n$ is additional information added to improve continuum convergence, and not intrinsic to the lattice model studied.
This means this choice of $q_n$ is not necessarily the sector of lowest energy for every given parameter choice of the lattice realization.

A transition between two ground-state sectors was identified early on by Steinhardt \cite{steinhardtSU2flavorSchwingerModel1977} in the study of the lattice $N_f=2$ flavor Schwinger model with equal mass. In a spin Hamiltonian obtained by a Jordan-Wigner transformation of the staggered 2-flavor model, he observed a phase transition from a ferromagnetic (Ising-type) ground-state to a Heisenberg-like ground-state. 
In particular, the lattice ground-state gauge sector changes from staggered $q_n=1-(-1)^n$ (Ising) to homogeneous $q_n=1$ (Heisenberg) background charges. 
One way to interpret this transition is that the gauge sector enforces only the sum of local background charges; if the mass is small enough, the lattice system favors shifting the staggering of the two flavors by half a unit cell. This effectively puts electrons of one flavor on the same site as the positrons of the other, allowing charge-bound states of the two flavors without the extra energy cost of a gauge link. Steinhardt also points out that the transition specific to the lattice theory, with the consequences unclear for the continuum theory.

\begin{figure}
    \centering
    \includegraphics{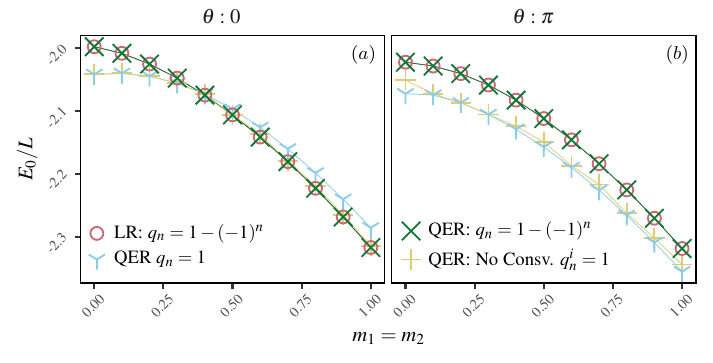}
    \caption{\label{fig:OpenSystemGaugeSector} Ground-state energy density of the $N_f=2$ Schwinger model with equal mass $m_1=m_2$ and open boundary conditions obtained by DMRG.  Our qubit encoded rotor (QER) with gauge conservation replicates the typical long range (LR) model (red circles) perfectly when the initial state is in the correct Gauss's Law sector $q_n = 1-(-1)^n$ (green crosses). If the gauge conservation is deactivated, the system initialized in the $q_n=1-(-1)^n$ sector is able to find lower energy solutions (yellow pluses) given the gap is big enough. In this case, this is the  $q_n=1$ (light blue Y) for small $m_1=m_2$ and $\theta=0$ or for all $m_1=m_2$ shown with $\theta=\pi$}
\end{figure}

This transition is seen in the DMRG results of Fig.~\ref{fig:OpenSystemGaugeSector} $(a)$. If the gauge constraint is enforced, the ground-state energy of the $q_n=1$ (blue tripoints) and the $q_n=1-(-1)^n$ sector cross.  The standard $q_n=1-(-1)^n$ sector is labeled by green crosses, while the red circles are an independent implementation using the long-range model without explicit gauge dynamics. If the constraint is relaxed, but initialized in the $q_n=1$ sector (yellow pluses), the variational method follows the points of lowest energy. We want to emphasize that all of these solutions satisfy the equation 
\begin{align}
    G_n\ket{\psi} = q_n \ket{\psi}
\end{align}
and are therefore gauge invariant. Only the background charge configuration $q_n$ changes.
The effect is even more severe in the presence of a $\theta=\pi$ term shown in Fig.~\ref{fig:OpenSystemGaugeSector}~$(b)$. Here, the $q_n=1$ sector is of lower energy for the whole commonly studied parameter space~\cite{Dempsey2024}, despite the fact that gauge invariance of the ground state is guaranteed for unconstrained simulations due to Elitzur's theorem.  This shows that variational approaches can and will converge to unphysical sectors if gauge charges are not explicitly conserved. These issues are avoided for arbitrary dimensions and or boundaries using the virtual rishon formulation. What follows is a discussion on the Schwinger model in the presence of open vs. periodic boundaries.

\subsection{Boundary effects in the lattice Schwinger model}
\begin{figure}
    \centering
    \includegraphics{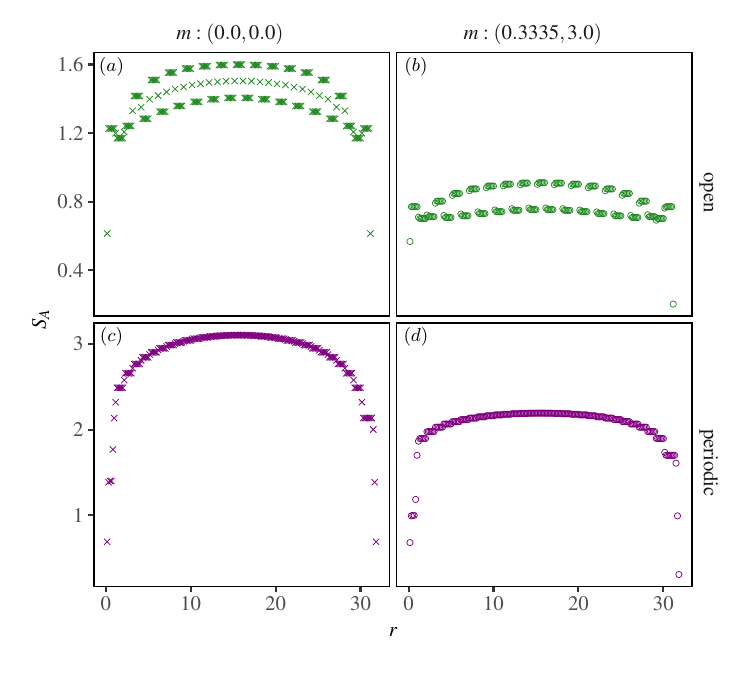}
    \vspace{-2em}
    \caption{\label{fig:BoundaryEffects} Boundary effects of the entanglement entropy in the Schwinger model for the example of the $N_f=2$ flavor case with $\theta=\pi$, $a=0.3$, $g=1$, and $L=32$. The open system (top row) shows severe boundary effects in form of staggering at the SU$(2)_1$ point $m_1=m_2=0$  (left column) and the Ising transition $m_1=0.3335, m_2=3$ (right column). This makes extraction using Eq. \eqref{eq:CardyFormula}, from a single wavefunction dependent on the branch chosen and therefore unreliable~\cite{ikedaDetectingCriticalPoint2023}. While more elaborate finite size entanglement ratios~\cite{Dempsey2024, campostriniFinitesizeScalingQuantum2014} with respect to the total system size are possible, this is not necessary for periodic boundary conditions.}
\end{figure}

At $\theta=\pi$ the lattice Schwinger model suffers from severe boundary effects. If studied for open systems, this not only takes form in physical observables like the electric field, but also in the entanglement entropy of bipartitions like the one shown in \ref{fig:CentralCharge} $(a)$. This is shown in the example of the critical points of the $N_f=2$ case in Figure \ref{fig:BoundaryEffects}. Focusing on the results of the open boundary systems in the top row, one identifies obvious splittings reminiscent of the staggering of the lattice formulation. For periodic boundary conditions (bottom half), on the other hand, this staggering effect vanishes completely. 

Open boundary conditions are much more commonly studied as these can be  simulated without a dynamic gauge field by employing a unitary mapping which introduces all-to-all couplings~\cite{banulsMassSpectrumSchwinger2013, Banuls:2016gid, Dempsey2024, papaefstathiouRealtimeScatteringLattice2025,ikedaDetectingCriticalPoint2023}. These boundary effects make the extraction of the central charge using Eq. \eqref{eq:CardyFormula} nontrivial. Single wavefunction scalings are nearly impossible but even system size scalings can give rise to the incorrectly identified central charges~\cite{ikedaDetectingCriticalPoint2023}. One way to avoid this is by the use of more elaborate entanglement-derived quantities~\cite{Dempsey2024, campostriniFinitesizeScalingQuantum2014}.

Periodic boundary conditions, on the other hand, completely alleviate the necessity of these elaborate finite size scalings, but require at least the simulation of one dynamic boson or a form of truncation~\cite{byrnesDensityMatrixRenormalization2002}. As the virtual rishon formulation faithfully simulates the gauge field, the study of periodic systems is the obvious choice for extracting the central charge. Of course, due to the increased bipartition boundary of periodic systems, the usual computational limitations of MPS when simulating periodic boundaries apply. This means that periodic systems require substantially higher total bond dimensions than an open system with similar parameters and system size. But due to the conservation of the extensive set of local charges, the computational resources in the folded ordering scale more favorably than usual. The wavefunction of a periodic system with equal total bond dimension has significantly smaller blocks, resulting in less memory requirements compared to the open wavefunction with the same total bond dimension. This is shown in Table \ref{tab:Periodic_memory_scaling}. Where an open system with bond dimension $\chi=2048$ can cost over 16GB RAM, while the periodic system requires only 9.7GB.

\begin{table}[h]
    \centering
    \caption{Total memory scaling (MPS+MPO+MPO-Environments) as logged by TeNPy for open and periodic boundary conditions, and the two different critical points of Figure \ref{fig:BoundaryEffects}. While periodic systems require much larger bond dimensions to achieve similar wavefunction accuracy, due to extensive charge conservation, the blocks in periodic systems are much smaller, thereby reducing the additional cost. }
    \label{tab:Periodic_memory_scaling}
    \begin{tabular}{lccrr}
        \toprule
        Panel & \textbf{$m_\alpha$} & bc & $\chi$ & Memory \\
        \midrule
        $(a)$&(0.0, 0.0)& open & 512 & 1619 MB\\
        $(a)$&(0.0, 0.0)& open & 1024 & 5120 MB\\
        $(a)$&(0.0, 0.0)& open & 2048 & 16324 MB\\
        $(b)$&(0.3335, 3.0) & open &512& 1418 MB\\
        $(b)$&(0.3335, 3.0)& open &804 & 7250 MB\\
        $(c)$&(0.0, 0.0)& periodic &512& 1084 MB\\
        $(c)$&(0.0, 0.0)& periodic &1024& 3025 MB\\
        $(c)$&(0.0, 0.0)& periodic &2048& 9773 MB\\
        $(d)$&(0.3335, 3.0) & periodic & 512& 1005 MB\\
        $(d)$&(0.3335, 3.0) & periodic & 1024& 2715 MB\\
        $(d)$&(0.3335, 3.0) & periodic & 2048& 8853 MB\\
        \bottomrule
    \end{tabular}
\end{table}
\subsection{\label{sec:disagreement_central_charge} Disagreement for central charges in high $m_3$ limit}
\begin{figure}
    \centering
    \includegraphics{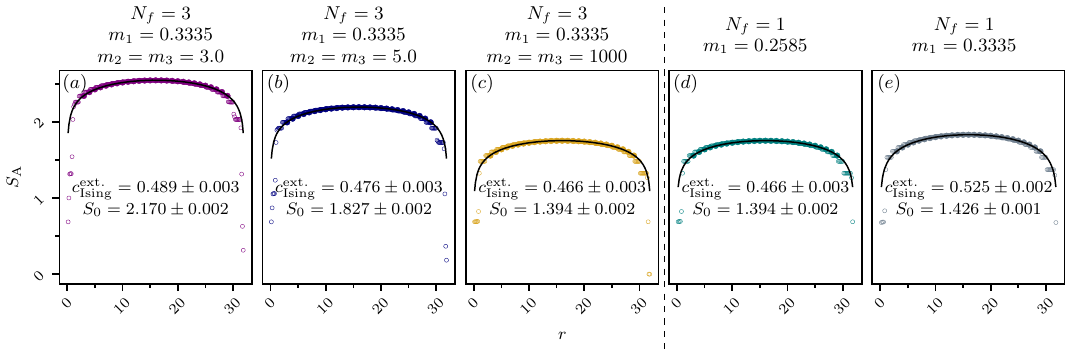}
    \vspace{-2em}
    \caption{\label{fig:IsingCentralChargeUnder} Underestimation of the Ising central charge for $N_f=3$ flavor Schwinger model. The Ising central charge shown in Fig.~\ref{fig:CentralCharge}~$(f)$ and here replotted in panel $(a)$, is underestimated with $c^{\rm ext.}_{\rm Ising} = 0.489 \pm 0.003$. At first sight, one expects an overestimation due to additional contributions from the modes suppressed by $m_{2} = m_3=3.0$; the small system size would also favor an overestimation (compare the $N_f=1$ equivalent in panel $(e)$). Further increasing the heavy modes to $m_{2} = m_3 = 5.0$, panel $(b)$, and $m_{2} = m_3=1000.0$, reveals a convergence towards the $N_f=1$ model with $m_1 = 0.2585$, panel $(d)$, missing the critical point. This is due to the hidden flavor number $N_f$ dependence of the mass-shift, $\tilde{m} =m - g^2 N_f /8a$. This shifts the simulated lattice mass by $ag^2(3-1)/8=0.075$ compared to the $N_f=1$ case. All results presented are as in Fig.~\ref{fig:CentralCharge}, using $a=0.3$, $\theta=\pi$, $g=1$, and  $N_r=4$ qubits for the rotor encoding.}
\end{figure}

Fig.~\ref{fig:CentralCharge} shows good agreement with the expected central charges. While the other central charges tend to overestimate the analytic results, the Ising critical point in the $N_f=3$ case, with $m_1=0.3335$ and $m_{2} = m_3 = 3.0$, is underestimated. This is shown in Fig~\ref{fig:CentralCharge}~$(f)$ and replotted with different y-scale in Fig.~\ref{fig:IsingCentralChargeUnder}~$(a)$. Here, the extracted central $c^{\rm ext.}_{\rm Ising} = 0.489 \pm 0.003$ has a relative error of about $2\%$. Importantly, numerical experiments of small lattice sizes would suggest an over vs. underestimation; together with the small but finite positive contributions to entanglement from the heavy fermionic modes with mass $m_2=m_3=3.0$, this makes the result for the central charge surprising at first. The deviation can however be explained by the hidden $N_f$ dependence of the corrected lattice mass used in this simulation \cite{Dempsey:2022nys, Dempsey2024},
\begin{align}
\tilde{m} = m - N_f\frac{g^2a}{8}    \,.
\end{align}
While this ensures the recovery of a discrete chiral symmetry in the vanishing-mass limit, in an anisotropic mass limit like the one discussed here, it introduces an $N_f$-dependent detuning of the light mass(es). 
For $g=1,a=0.3, \theta=\pi$, the $N_f=3$ model with $m_1 = 3335$ and $m_{2}, m_3 \to \infty$ limit should match the simulation of the lattice model with $N_f=1$ and $m_1=0.3335-(3-1)\frac{ag^2}{2}= 0.2585$, therefore slightly missing the critical point independent of finite system size $L$ corrections to its position.

This is confirmed by numerical data provided in \ref{fig:IsingCentralChargeUnder}. Here the we show that the extracted central charge reduces further when approaching the $m_{2}, m_3 \to \infty$ limit following panel $(a)$ with $m_{2} = m_3 =3.0$ over $(b)$ with $m_{2} = m_3 =5.0$ to $(c)$ with $m_{2} = m_3=1000$. In the extreme limit the $N_f=3$ model matches the $N_f=1$, $m_1=0.2585$ (panel $(d)$) with  $c^{\rm ext.}_{\rm Ising} = 0.466 \pm 0.003$ compared to the $N_f=1$ at $m_1=0.3335$ in panel $(e)$, which shows the expected overestimation of the central charge ($c^{\rm ext.}_{\rm Ising} = 0.525 \pm 0.002$) due to the relatively small lattice size of $L=32$.

\subsection{\label{sec:ConvergenceAnalysisSchwinger}Convergence analysis of the Schwinger ground-state data}

\begin{figure}
    \centering
     \includegraphics{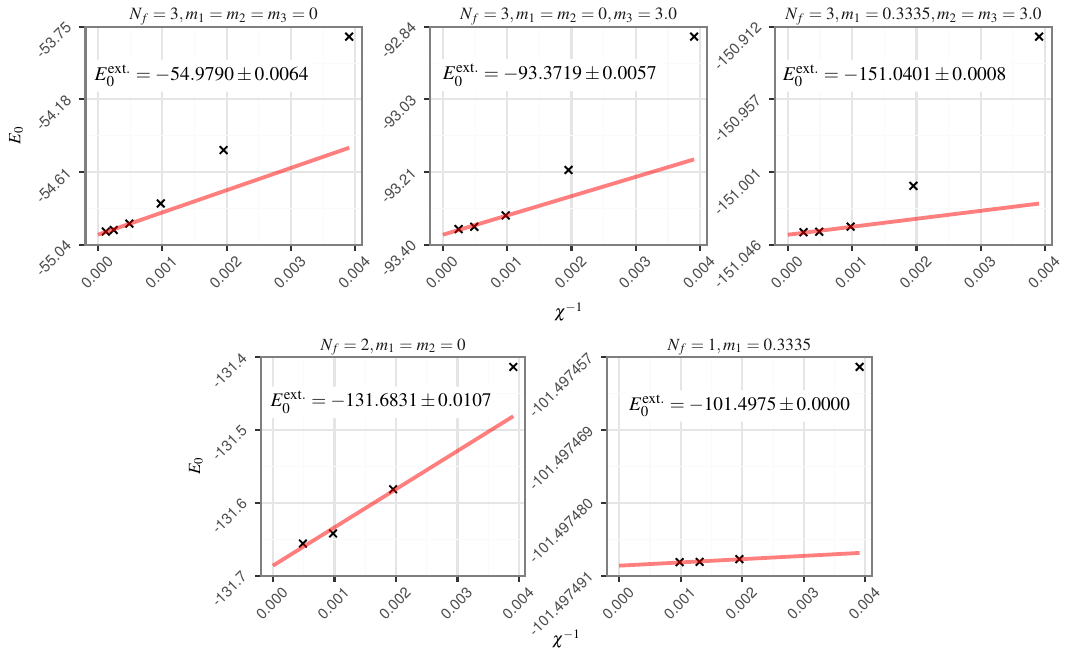}
        \caption{Energy convergence $E_0$ of the Schwinger ground-state~(shown in Fig.~\ref{fig:CentralCharge}) with respect to the inverse of the maximum MPS bond dimension $\chi$. The number of free parameters in a MPS wavefunction scales with $\chi^2$. As expected for the variational method DMRG, the energy decreases monotonically and converges to a constant as the limit $\chi \rightarrow \infty$ or $\chi^{-1} \to 0$ is approached. In the respective figures, we show the extrapolated infinite $\chi$ limit  $E_0^{\rm ext.}$ obtained from performing a linear fit (red line) of the highest three bond dimensions for each system simulated.  The uncertainties are obtained from these 3 data points and should be understood as estimates. The entanglement entropies depicted in Fig.~\ref{fig:CentralCharge} are from the wavefunctions matching the data-points of highest $\chi$ shown here. These extrapolated energies are not used anywhere else in the paper.}
        \label{fig:convergenceground-stateSchwinger}
\end{figure}
The SU$(2)_1$ and SU(3)$_1$ critical points with periodic boundary conditions, are numerically taxing due to their high entanglement reflected by the prefactor of their area law correction $\frac{c}{3}\log(\frac{L}{\pi}\sin(\pi \frac{r}{L}))$. To ensure simulation accuracy while reducing finite-size effects, we opted for the strongest bond dimension since the  
accuracy of matrix product state simulation is governed by the MPS bond dimension. DMRG, as a variational method, is expected to converge towards the true ground-state as the number of variational parameters is increased via the bond dimensions $\chi$. With the convergence of the energy $E_0$ as a measure of accuracy, in this work, we used the following DMRG protocol:
\begin{enumerate}
    \item Run DMRG
    \begin{itemize}
        \item Initiate subspace expansion mixer with high amplitude $0.9$ and exponential decay during the sweeps.
        \item Set maximum allowed bond-dimension $\chi_{max,i}$
        \item Run two site DMRG until we reach $\delta E_0 \leq 10^{-10}$ and $\delta S \leq 10^{-7}$ or we reach the max number of sweeps $=50$.
    \end{itemize}
    \item once convergence is reached, we increase the $\chi_{max,i}$ and use the last MPS wavefunction as the inital guess
    \item Repeat until $\chi_{max, i} > \chi_{max}$
\end{enumerate}
For each intermediate result $\chi_{max, i}$, we calculate the ground-state energy, entanglement entropy, etc. 
While we only show the entropy for the highest performed bond-dimension in Fig.~\ref {fig:CentralCharge}, we show here the ground-state energy with respect to the inverse intermediate $\chi$ in Fig. \ref{fig:convergenceground-stateSchwinger}. Generally, extrapolating the energy with respect to the truncation error is preferred, but in our case, the gauge-constrained system often yields $\epsilon_{trunc} = 0$ once the mixer is deactivated, as in single-site DMRG. Due to this, we resort to the empirical alternative of performing a linear fit of the inverse of the highest 3 $\chi$ values to extrapolate towards the $\chi^{-1} \rightarrow 0$ limit and get an approximation of the energy $E^{\rm ext.}_{0}$. The deviation of the last ground-state energy from this extrapolated point and its uncertainty should be understood as an estimate of the simulation's convergence.

\section{Details on the results of QED$_3$}
\begin{figure}
    \centering
    \includegraphics{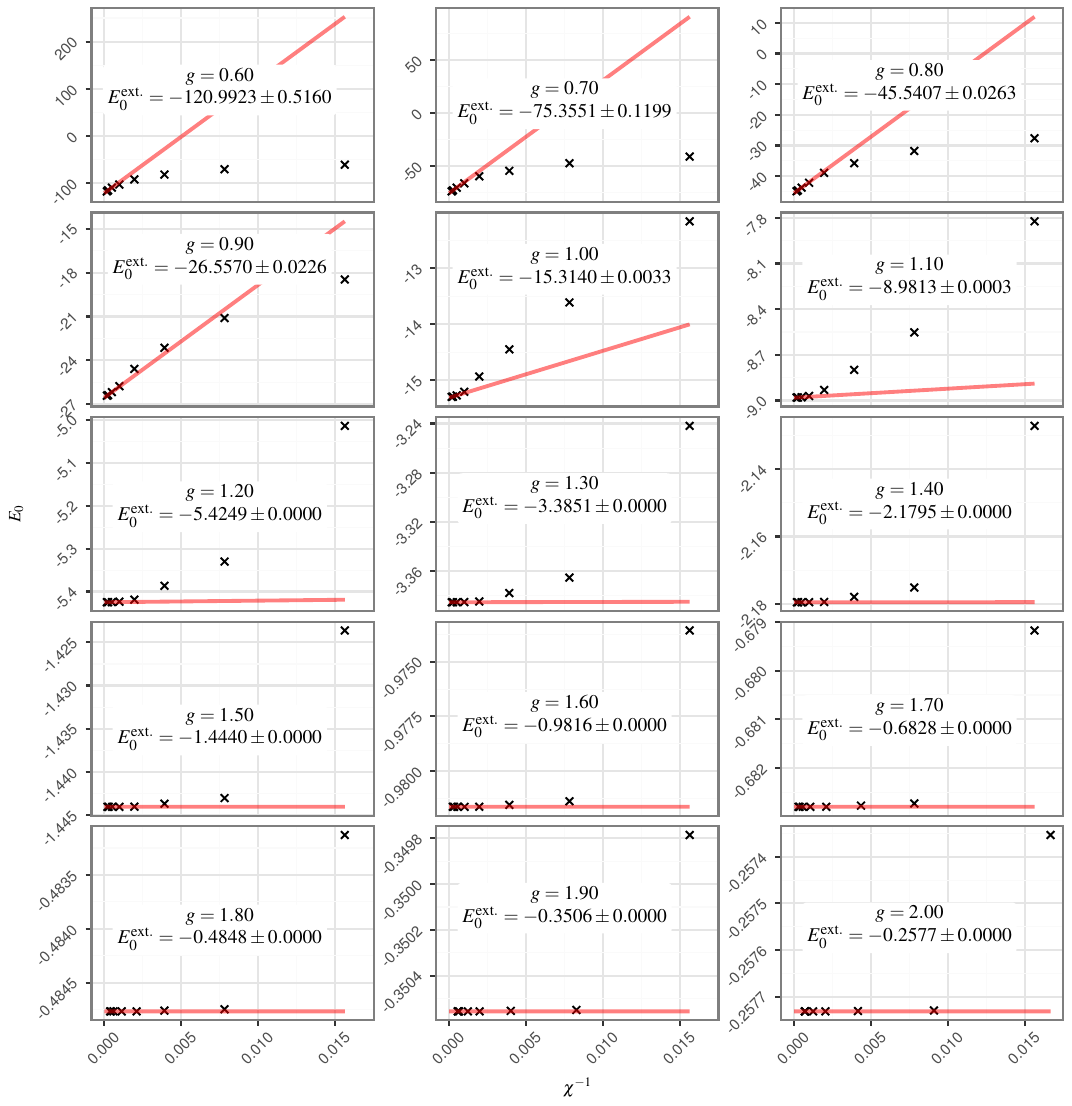}
    \caption{ground-state energy convergence $E_0$ with respect to the inverse of the maximum bond-dimension $\chi$ for QED$_3$ without background charges. With the increase of bond dimensions $\chi$, the energy converges monotonically to a constant. While clear convergence can be observed for $g>1$, an increasing slope (red curve) obtained from the highest 3 $\chi$ values results in stronger and stronger uncertainties in the ground-state energy extrapolation $E_0^{\rm ext.}$. This indicates that bond dimensions beyond those presented here, up to $\chi_{\rm max} = 6144$, are required to accurately capture the wavefunction in these parameter regimes. While only the background charge less case $\Delta=0$ is shown here, similar results are obtained for $\Delta\neq 0$, and each extracted value is used in \ref{fig:StringTensionFites}, to obtain the string tension $\sigma$ shown in \ref{fig:QED3} $(d-e)$.}
    \label{fig:convergenceground-stateQED}
\end{figure}

In this section, we describe the computation of the string tension $\sigma$.
Each string tension data point shown in Fig.~\ref{fig:QED3}~$(d-e)$ was obtained as follows:
\begin{itemize}
    \item For each $g$ and $\Delta$, perform an independent DMRG calculation as described in Sec. \ref{sec:ConvergenceAnalysisSchwinger}.
    \item Extrapolate $E_0$ towards $\chi \rightarrow\infty$ by performing a linear fit of $\chi^{-1} \propto E_0 $ for $\chi \in [2048, 4096, 6144]$ and use the fit error as uncertainty estimate.
    \item Perform an weighted linear fit of $E_0^{\rm ext.} \propto \Delta $ to extract $\sigma$ for $\Delta \in [1,8]$.
\end{itemize}
The infinite $\chi$ extrapolation is shown only for the case without any string excitation $\Delta=0$, in Fig~\ref{fig:convergenceground-stateQED}. While the value for $g > 1$ shows clear convergence with nearly vanishing slope and uncertainties below $10^{-4}$, the data still shows a clear slope for $g < 1$ with increasing uncertainties. This reveals that the low $g$ regime is challenging in the MPS simulation used here. Results for other $\Delta\neq0$ are obtained similarly and are qualitatively the same. These are therefore not shown here.

\begin{figure}
    \centering
    \includegraphics{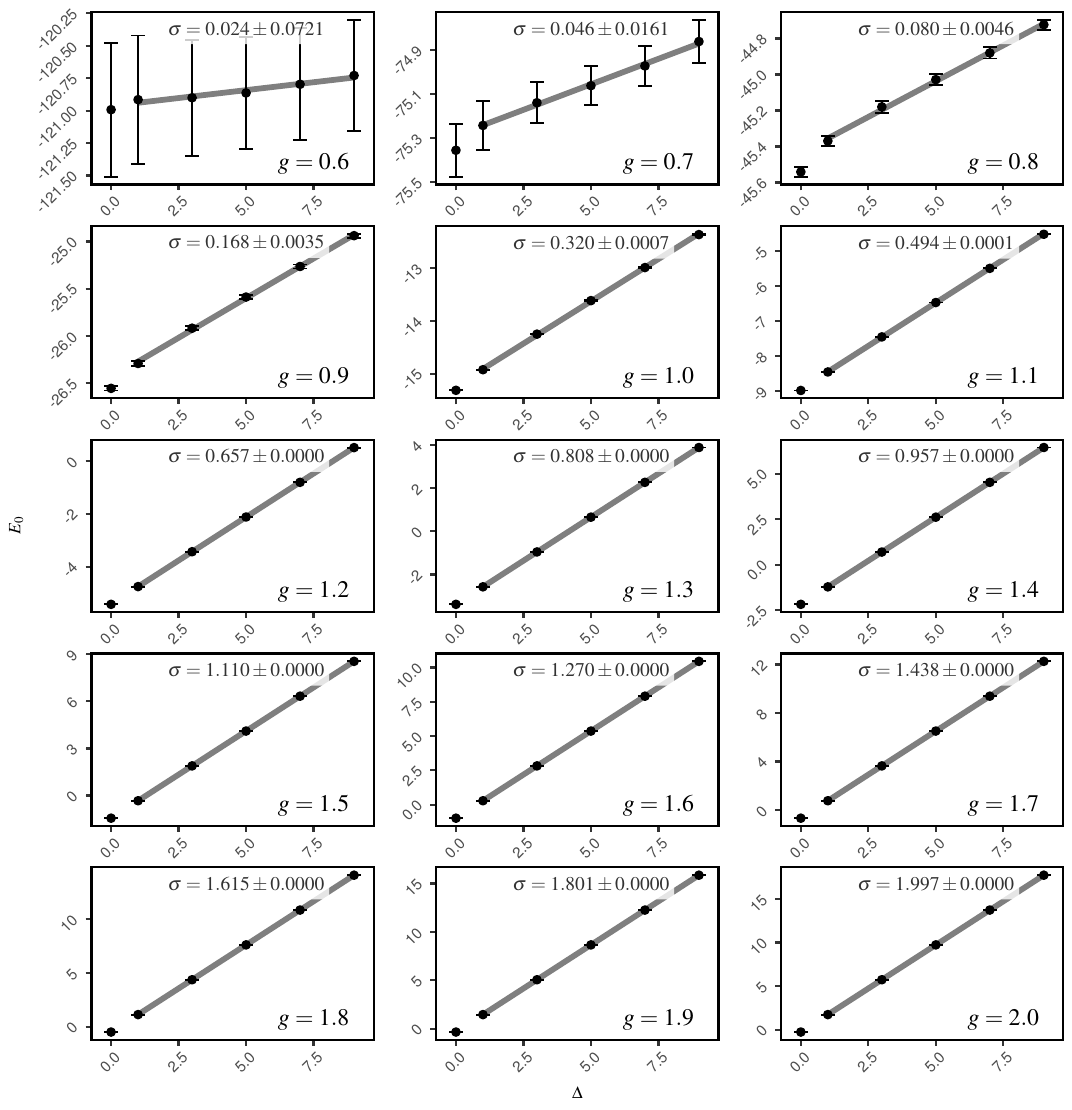}
    \vspace{-2em}
    \caption{\label{fig:StringTensionFites} ground-state energy $E_0$ vs length of the string excitation $\Delta$ for $g \in [0.6, 2.0]$ in ${\rm QED}_3$. The ground-state energy increases proportionally with charge separation $\Delta$ with the constant of proportionality $\sigma$ being the string tension. The computational limitations at $g\rightarrow0$ lead to significant uncertainties in the extraction of the energy, as reflected in the error bars shown at one-sigma confidence. The values extracted here are shown as data-points in Fig.~\ref{fig:QED3} $(c-d)$. As described in the main text, each point is an individual DMRG simulation, with $a=1$, $L_y=7, L_x=12$, and $N_r=4$ qubits per rotor.}
\end{figure}

These extracted interpolated values are then used to extract the string tension for each $g$ as shown in Fig~\ref{fig:StringTensionFites}. Each point is the interpolated energy, with the fit error (square root of the respective entry of the covariance matrix) used as the uncertainty of the error bars. The weighted linear fits result in the $\sigma$ values shown.

\section{Data and source code availability}
The simulation's source code is based on the TeNPy tensor network library~\cite{hauschildTensorNetworkPython2024}. The data analysis was performed using standard python libraries: Numpy, Pandas, Plotnine. The data is structured using signac~\cite{ramasubramaniSignacPythonFramework2018}. A repository able to reproduce the results shown in this publication is provided under \cite{rogerson_2026_18869792}.

\subsection{Data availability}
All data used in this letter is provided in a Zenodo data repository~\cite{rogerson_2026_18864580}. With signac, each run is defined by a set of parameters in a JSON file, from which a unique job ID (hash) is generated. The JSON parameter file, as well as all direct results (in HDF5 format) from each simulation run, are stored in their respective folders named after the ID. While this makes the folder structure less convenient for human readability, it allows fast querying via the signac python library. For an in-depth analysis of the data, we suggest installing signac to query the data. For a quick overview, we provide the first few digits of the corresponding Job IDs of each figure in the paper in the following tables.
\begin{table}[h]
\centering
\caption{Mapping of figure panels to specific simulation data Ids.}
\label{tab:sim_mapping}
\begin{tabular}{l|lll||l}
\toprule
\textbf{Figure} & \textbf{Panel} & \textbf{Marker} & \textbf{Color} & \textbf{ID (Hash)} \\
\midrule
\ref{fig:CentralCharge} & (b) &Circle ($\circ$) & Forestgreen  & \texttt{d44327\dots} \\
\ref{fig:CentralCharge} & (c) &Circle ($\circ$) & Purple  & \texttt{daac83\dots} \\
\ref{fig:CentralCharge} & (d) &Circle ($\circ$) & Maroon  & \texttt{705bd3\dots} \\
\ref{fig:CentralCharge} & (e) &Circle ($\circ$) & Forestgreen  & \texttt{fa6595\dots} \\
\ref{fig:CentralCharge} & (f) &Circle ($\circ$) & Purple  & \texttt{ed3120\dots} \\
\bottomrule
\ref{fig:IsingCentralChargeUnder} & (a) &Circle ($\circ$) & Purple  & \texttt{ed3120\dots} \\
\ref{fig:IsingCentralChargeUnder} & (b) &Circle ($\circ$) & Navy  & \texttt{133843\dots} \\
\ref{fig:IsingCentralChargeUnder} & (c) &Circle ($\circ$) & Yellow  & \texttt{c1ee62\dots} \\
\ref{fig:IsingCentralChargeUnder} & (d) &Circle ($\circ$) & Teal  & \texttt{c0d149\dots} \\
\ref{fig:IsingCentralChargeUnder} & (e) &Circle ($\circ$) & Grey  & \texttt{fb1357\dots} \\
\bottomrule
\ref{fig:BoundaryEffects} & (a) & Cross ($\times$) & Forestgreen  & \texttt{6de3e2\dots} \\
\ref{fig:BoundaryEffects} & (b) & Circle ($\circ$) & Forestgreen & \texttt{aa9e2a\dots} \\
\ref{fig:BoundaryEffects} & (c) & Cross ($\times$) & Darkpurple & \texttt{d48d6b\dots} \\
\ref{fig:BoundaryEffects} & (d) & Cricle ($\circ$) & Darkpruple & \texttt{97959a\dots} \\
\bottomrule
\end{tabular}
\end{table}
\begin{table}
    \caption{Job IDs for all data points shown in  Fig. \ref{fig:OpenSystemGaugeSector}}
    \label{tab:JobIdsOpenSystemGaugeSector}
    \begin{tabular}{c|cccc||cccc}
        \toprule
        & \multicolumn{4}{c||}{$\theta=0$} & \multicolumn{4}{c}{$\theta=\pi$} \\
        \midrule
        $m_1 = m_2$& \rotatebox{90}{LR: $q_n = 1-(-1)^n$} & \rotatebox{90}{QER $q_n = 1$} & \rotatebox{90}{QER: $q_n = 1-(-1)^n$} & \rotatebox{90}{QER: No Consv. $q_n^i = 1$} & \rotatebox{90}{LR: $q_n = 1-(-1)^n$} & \rotatebox{90}{QER $q_n = 1$} & \rotatebox{90}{QER: $q_n = 1-(-1)^n$} & \rotatebox{90}{QER: No Consv. $q_n^i = 1$} \\
        \midrule
        0.0 & \texttt{541d56}\dots & \texttt{798446}\dots & \texttt{644231}\dots & \texttt{6372f5}\dots & \texttt{c23443}\dots & \texttt{c1fe62}\dots & \texttt{f4e233}\dots & \texttt{74f822}\dots \\
        0.1 & \texttt{5c09c6}\dots & \texttt{d80557}\dots & \texttt{f11c08}\dots & \texttt{141c85}\dots & \texttt{b993fd}\dots & \texttt{2d8998}\dots & \texttt{82183d}\dots & \texttt{4fc032}\dots \\
        0.2 & \texttt{3b2f3f}\dots & \texttt{d1914a}\dots & \texttt{01d7ef}\dots & \texttt{5b0481}\dots & \texttt{ebc3d1}\dots & \texttt{98b191}\dots & \texttt{1d8ce0}\dots & \texttt{67ca11}\dots \\
        0.3 & \texttt{461539}\dots & \texttt{d3a62f}\dots & \texttt{e40eb5}\dots & \texttt{be7d5a}\dots & \texttt{efdfbb}\dots & \texttt{a208fa}\dots & \texttt{67fa40}\dots & \texttt{299420}\dots \\
        0.4 & \texttt{4cccc6}\dots & \texttt{dc4be0}\dots & \texttt{e608ac}\dots & \texttt{48ba36}\dots & \texttt{f47c7d}\dots & \texttt{e6d371}\dots & \texttt{b8c95e}\dots & \texttt{800a60}\dots \\
        0.5 & \texttt{031b0c}\dots & \texttt{6420b3}\dots & \texttt{1b1777}\dots & \texttt{ed56c2}\dots & \texttt{3ace6d}\dots & \texttt{1aa73d}\dots & \texttt{1047cb}\dots & \texttt{c9f55d}\dots \\
        0.6 & \texttt{19e3e2}\dots & \texttt{29072a}\dots & \texttt{c7ae34}\dots & \texttt{41faff}\dots & \texttt{f4b1bd}\dots & \texttt{8c548e}\dots & \texttt{4dc3f2}\dots & \texttt{334248}\dots \\
        0.7 & \texttt{96bfd9}\dots & \texttt{547ded}\dots & \texttt{8a3810}\dots & \texttt{af3ba2}\dots & \texttt{9e2b1b}\dots & \texttt{5f4f85}\dots & \texttt{6f31d1}\dots & \texttt{cd1fd7}\dots \\
        0.8 & \texttt{c48538}\dots & \texttt{1be671}\dots & \texttt{3dbe09}\dots & \texttt{7b703f}\dots & \texttt{ba9aaf}\dots & \texttt{0396a1}\dots & \texttt{d6d506}\dots & \texttt{5887f6}\dots \\
        0.9 & \texttt{b061da}\dots & \texttt{a3ab18}\dots & \texttt{03c777}\dots & \texttt{eefd88}\dots & \texttt{57ce7d}\dots & \texttt{0c5132}\dots & \texttt{f4aee2}\dots & \texttt{c79d4d}\dots \\
        1.0 & \texttt{9f8a6d}\dots & \texttt{f03016}\dots & \texttt{c0fdd2}\dots & \texttt{44ee8b}\dots & \texttt{97e437}\dots & \texttt{345ec8}\dots & \texttt{3fbfb6}\dots & \texttt{3dab02}\dots \\
        \bottomrule
    \end{tabular}
\end{table}

\begin{table}
\caption{Job IDs for the QED$_3$ data shown in this work. $\Delta=5$ and $g \in \{0.8, 1.2, 1.6\}$ are used in Fig. \ref{fig:QED3} (a)-(c).  The $\Delta=0$ subset is used in Fig. \ref{fig:convergenceground-stateQED}. All IDs are used in Fig.~\ref{fig:StringTensionFites} and Fig.~\ref{fig:QED3} (d), (e).}
\label{tab:JobIdsQED3}
\begin{tabular}{l|llllll}
\toprule
g & $\Delta=0$ & $\Delta=1$ & $\Delta=3$ & $\Delta=5$ & $\Delta=7$ & $\Delta=9$ \\
\midrule
0.6 & \texttt{05f500}\dots & \texttt{36ae1d}\dots & \texttt{7852ea}\dots & \texttt{15fa70}\dots & \texttt{353d33}\dots & \texttt{4bdc49}\dots \\
0.7 & \texttt{dbccb9}\dots & \texttt{3a7445}\dots & \texttt{c4c407}\dots & \texttt{b06355}\dots & \texttt{5fd8dd}\dots & \texttt{09abda}\dots \\
0.8 & \texttt{d0dbd1}\dots & \texttt{5277c8}\dots & \texttt{75a75b}\dots & \texttt{11ce80}\dots & \texttt{5ce6eb}\dots & \texttt{a33cad}\dots \\
0.9 & \texttt{728883}\dots & \texttt{af8b00}\dots & \texttt{372d82}\dots & \texttt{ea75aa}\dots & \texttt{4d504f}\dots & \texttt{aa6d8e}\dots \\
1.0 & \texttt{f86374}\dots & \texttt{15dcea}\dots & \texttt{a7a7a4}\dots & \texttt{820ac8}\dots & \texttt{c870b2}\dots & \texttt{700feb}\dots \\
1.1 & \texttt{610c81}\dots & \texttt{af8891}\dots & \texttt{a6c9bf}\dots & \texttt{19a444}\dots & \texttt{a19e23}\dots & \texttt{a30758}\dots \\
1.2 & \texttt{cc246a}\dots & \texttt{ebdd3a}\dots & \texttt{2ad88f}\dots & \texttt{3255cd}\dots & \texttt{3baca2}\dots & \texttt{46a173}\dots \\
1.3 & \texttt{022822}\dots & \texttt{d535cd}\dots & \texttt{32fd0b}\dots & \texttt{5a37b9}\dots & \texttt{dfe41a}\dots & \texttt{f3719a}\dots \\
1.4 & \texttt{cd9299}\dots & \texttt{cb0078}\dots & \texttt{74c343}\dots & \texttt{d8915d}\dots & \texttt{0f9a37}\dots & \texttt{eaa622}\dots \\
1.5 & \texttt{3b662d}\dots & \texttt{a6da83}\dots & \texttt{d6b5d0}\dots & \texttt{85798b}\dots & \texttt{0281d2}\dots & \texttt{9573b6}\dots \\
1.6 & \texttt{3bc342}\dots & \texttt{b1b41e}\dots & \texttt{a28a87}\dots & \texttt{aede5a}\dots & \texttt{6013c9}\dots & \texttt{3e2420}\dots \\
1.7 & \texttt{c206f6}\dots & \texttt{37e506}\dots & \texttt{210899}\dots & \texttt{7086b5}\dots & \texttt{cc2c7d}\dots & \texttt{28f660}\dots \\
1.8 & \texttt{034290}\dots & \texttt{0aba0b}\dots & \texttt{d2ee19}\dots & \texttt{e215ee}\dots & \texttt{5df8b5}\dots & \texttt{ac1321}\dots \\
1.9 & \texttt{757835}\dots & \texttt{2e9f31}\dots & \texttt{eac2d0}\dots & \texttt{5341d8}\dots & \texttt{0e3054}\dots & \texttt{19e938}\dots \\
2.0 & \texttt{066df6}\dots & \texttt{a91f2d}\dots & \texttt{aad27c}\dots & \texttt{a7d730}\dots & \texttt{6d810b}\dots & \texttt{81b1a6}\dots \\
\bottomrule
\end{tabular}
\end{table}

\end{document}